\documentclass{article}

\usepackage{arxiv}
\usepackage{natbib}
\usepackage[utf8]{inputenc} 
\usepackage[T1]{fontenc}    
\usepackage{url}            
\usepackage{booktabs}       
\usepackage{amsfonts}       
\usepackage{nicefrac}       
\usepackage{microtype}      
\usepackage{lipsum}
\usepackage{graphicx}
\usepackage{amsmath}
\usepackage{float}
\usepackage{mathtools}
\usepackage{amssymb}
\usepackage{algorithm}
\usepackage{algpseudocode}
\usepackage{amsfonts}

\usepackage{multirow}

\DeclareMathOperator*{\argmin}{arg\,min}
\usepackage{tabularx}

\title{A Parametric Multiscale Surrogate Framework Based on Texture-Generalizable Deep Material Networks for Polycrystal Modeling}

\author{
 Ting-Ju Wei \\
  Department of Civil Engineering\\
  National Taiwan University\\
  Taipei, Taiwan \\
   \And
 Tung-Huan Su\thanks{Corresponding author. Email: \texttt{michsu@synopsys.com}} \\
  Synopsys Inc.\\
  Livermore, CA, USA \\
   \And
 Chuin-Shan Chen\thanks{Corresponding author. Email: \texttt{dchen@ntu.edu.tw}} \\
  Department of Civil Engineering\\
  Department of Materials Science and Engineering\\
  National Taiwan University\\
  Taipei, Taiwan \\
}

\begin{document}
\maketitle
\begin{abstract}
This work presents a computational framework for parametric multiscale surrogate modeling of polycrystalline materials. The framework integrates a physics-based Deep Material Network (DMN), specifically an Orientation-aware interaction-based Deep Material Network (ODMN), with two data-driven components: a Texture-Adaptive Clustering and Sampling (TACS) scheme that provides a reduced yet statistically consistent representation of crystallographic texture, and a Graph Neural Network (GNN) that infers the micromechanical equilibrium parameters of the ODMN from grain-level interaction graphs. Referred to as the TACS–GNN–ODMN framework, this combination introduces a microstructure-to-parameter mapping that constructs fully parameterized surrogate models for previously unseen microstructures without retraining. The resulting surrogate model preserves the micromechanical structure of the underlying formulation while substantially reducing computational cost. Numerical results show that the framework accurately predicts nonlinear mechanical responses and crystallographic texture evolution under several loading conditions, in close agreement with full-field direct numerical simulations (DNS). The method further achieves more than two orders of magnitude speed-up over fast Fourier transform (FFT)-based simulations. The framework thus provides an efficient and physically consistent strategy for multiscale modeling of polycrystalline materials and is well suited to large-scale simulations and real-time applications.
\end{abstract}


\section{Introduction}\label{sec1}

Accurate prediction of mechanical response and texture evolution across diverse polycrystalline microstructures remains a central challenge in multiscale materials modeling and process design. Crystal plasticity (CP) provides a fundamental framework for capturing intragranular deformation heterogeneity and predicting texture evolution under various loading conditions~\cite{ZHOU201619, KNEZEVIC2014239}. To bridge the micro-to-macro transition, a range of homogenization models has been developed. The classical Taylor model assumes uniform deformation across grains, often leading to overestimated flow stresses and exaggerated texture sensitivity. The self-consistent model improves on this by embedding each grain in an effective medium, yet it still fails to resolve local field fluctuations and grain interactions. The relaxed grain cluster (RGC) model partially addresses this limitation~\cite{tjahjanto2015multiscale}, but its reliance on empirically fitted parameters limits robustness and transferability.

Full-field CP methods overcome these limitations by discretizing the representative volume element (RVE) and directly solving the governing equations using finite element (FEM) or fast Fourier transform (FFT)-based solvers~\cite{roters2019damask, eisenlohr2013spectral, shanthraj2015numerically, lebensohn2020spectral}. While these approaches provide high-fidelity predictions and detailed insight into microstructural mechanisms~\cite{anand2004single, ardeljan2014dislocation, knezevic2010deformation}, their computational cost remains prohibitive for large-scale and real-time applications.

To alleviate this limitation, data-driven surrogate models have been developed to learn mappings from microstructures to macroscopic responses~\cite{su2022multiscale}. Recent approaches include neural-network-based constitutive models~\cite{SUN2021102973, IBRAGIMOVA2021103059, IBRAGIMOVA2022103374}, physics-constrained learning frameworks~\cite{ZHOU2024115861}, recurrent architectures for history-dependent behavior~\cite{BONATTI2022104697, BONATTI2022103430}, graph neural networks (GNNs) that explicitly encode grain-level interactions~\cite{hu2024temporal}, and parametrically upscaled crystal plasticity models~\cite{WEBER2022115384}. Although these methods demonstrate strong predictive capability, most existing approaches focus on direct response prediction, with limited emphasis on constructing an explicit mapping to a physics-based surrogate representation. As a result, the connection between the learned representation and the underlying micromechanical structure may be less explicit in multiscale settings.

Deep Material Networks (DMNs) address this gap by introducing a physics-informed hierarchical representation in which micromechanical equilibrium is enforced at every level~\cite{LIU201920, gajek2020micromechanics, noels2022micromechanics}. Trained on linear-elastic stiffness data, DMNs encode intrinsic geometry--mechanics relationships through their network architecture, while analytical upscaling and downscaling operators ensure equilibrium and compatibility during inference~\cite{wan2024decoding, WeiOverview, BHATKEELANJESRINIVAS2026118517}. This framework has been extended to a broad range of material systems and constitutive settings, including woven composites~\cite{shin2024deep2}, porous materials~\cite{noels2022interaction}, rigid-fiber suspensions in non-Newtonian fluids~\cite{sterr2025deep}, and composites with micropolar constitutive behavior~\cite{FRANCIS2025118329}. Recent studies have further explored transferability and parametric generalization strategies based on interpolation of DMN parameters~\cite{LI2024116687, huang2022microstructure}, as well as feature extractors based on GNNs~\cite{jean2024graph}, convolutional neural networks~\cite{WU2026118554}, and foundation models~\cite{wei2025foundation}. These developments establish DMNs as an efficient and physically consistent surrogate framework for nonlinear multiscale simulation.

Extending DMNs to polycrystals, the Orientation-aware interaction-based Deep Material Network (ODMN) incorporates crystallographic orientation into the micromechanical hierarchy, enabling prediction of both stress--strain responses and texture evolution~\cite{Wei01}. However, ODMN lacks a parametric mapping from arbitrary microstructures to its parameter space, which requires retraining for each new microstructure and limits transferability. This limitation is particularly important in applications with large numbers of diverse microstructures, where repeated retraining is computationally impractical.

To overcome this limitation, we reformulate ODMN generalization as a microstructure-to-parameter inference problem. We establish a parametric mapping from polycrystalline microstructures to the ODMN parameter space and propose a computational surrogate framework based on TACS--GNN--ODMN. In contrast to existing approaches that directly predict responses, the proposed method constructs a physics-based surrogate through an explicit microstructure-to-parameter mapping, thereby preserving micromechanical consistency and interpretability. The framework decomposes the problem into two components: a Texture-Adaptive Clustering and Sampling (TACS)~\cite{wanni2024machine} scheme for reduced yet statistically consistent texture representation, and a Graph Neural Network (GNN) for inferring micromechanical equilibrium parameters from grain-level interaction graphs. This formulation enables the construction of ODMNs for previously unseen microstructures without retraining while maintaining physical consistency.

The main contributions of this work are as follows. First, a parametric mapping from microstructures to ODMN representations is established, enabling generalization without retraining. Second, ODMN generalization is reformulated as a microstructure-to-parameter inference problem. Third, the proposed TACS--GNN--ODMN framework integrates texture representation and micromechanical parameter inference within a unified computational framework. Finally, the framework accurately predicts nonlinear mechanical responses and texture evolution for previously unseen microstructures in close agreement with direct numerical simulations (DNS).

The remainder of this paper is organized as follows. Section~\ref{sec2} introduces the TACS--GNN--ODMN framework. Section~\ref{sec3} presents numerical results. Section~\ref{sec4} concludes the paper.

\section{TACS--GNN--ODMN Framework}\label{sec2}

\begin{figure}[htbp]
    \centering
    \includegraphics[width=1\linewidth]{ 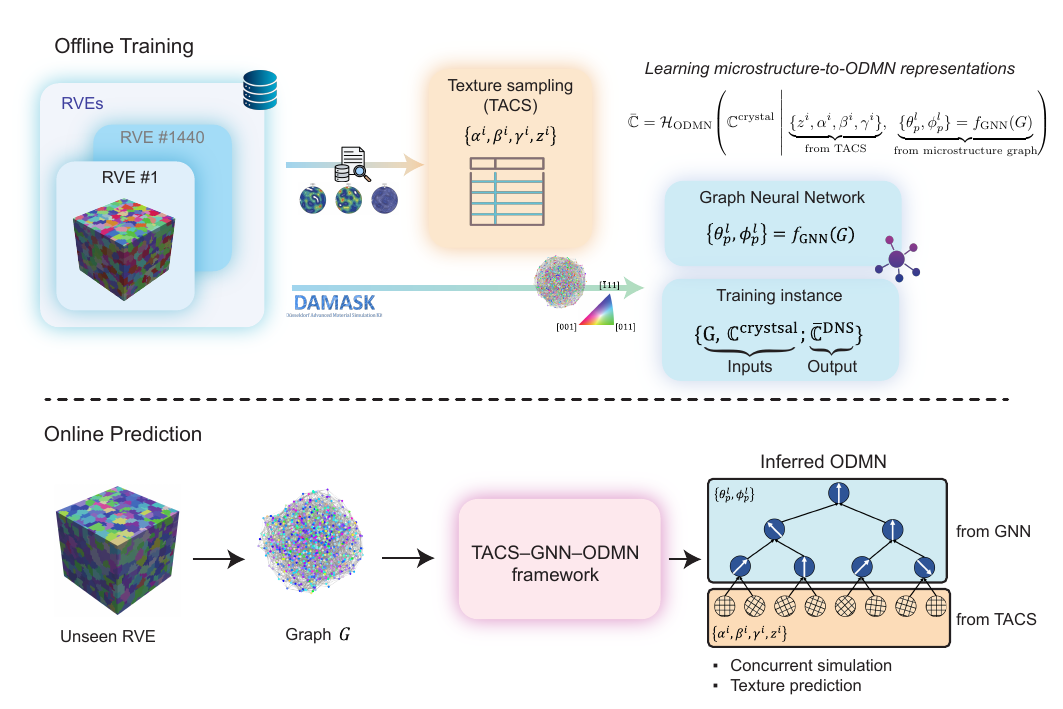}
    \caption{Schematic of the TACS--GNN--ODMN framework for microstructure-to-parameter inference. 
    In the offline stage, RVEs are converted into grain-level graphs, from which a GNN learns micromechanical equilibrium parameters, while the TACS scheme provides consistent initialization of texture-related parameters. 
    In the online stage, an unseen microstructure is encoded as a graph, and the trained GNN infers the corresponding ODMN parameters, which are combined with TACS-derived texture parameters to construct a fully parameterized ODMN. 
    This enables efficient prediction of homogenized responses and texture evolution without retraining.}
    \label{fig:framework}
\end{figure}

This section presents the TACS--GNN--ODMN framework, as illustrated in Fig.~\ref{fig:framework}. The framework formulates the construction of a parametric ODMN as a microstructure-to-parameter inference problem and operates in two stages: offline training and online prediction.

During offline training, the TACS scheme generates sampled Tait–Bryan angles that reconstruct the orientation distribution function (ODF), providing consistent initialization for the texture-related parameters of the ODMN. At the same time, the GNN learns a mapping from grain-level microstructure graphs to the parameters governing hierarchical micromechanical equilibrium in the ODMN. In these graphs, nodes encode crystallographic orientation, volume fraction, and geometric attributes, while edges represent topological adjacency. The combined system is trained end-to-end so that the parametric ODMN learns the homogenization mapping from constituent elastic stiffness matrices to homogenized responses, with ground truth obtained from full-field DNS.

During online prediction, an unseen RVE is converted into a grain-level graph, and the trained GNN infers the corresponding micromechanical equilibrium parameters. These parameters are combined with the TACS-initialized texture parameters to construct a fully parameterized ODMN tailored to the given microstructure. The resulting surrogate model then predicts homogenized mechanical responses and the evolution of crystallographic texture under arbitrary loading paths.

The following subsections describe each component of the framework. Section~\ref{sec:odmn} reviews the ODMN formulation. Section~\ref{sec:tacs} introduces the TACS initialization strategy. Section~\ref{sec:graph} presents the construction of microstructure graphs. Section~\ref{sec:gnn} describes the GNN architecture. Section~\ref{sec:dataset} describes dataset generation and ground-truth computation. Section~\ref{sec:offline} outlines the offline training procedure, and Section~\ref{sec:online} describes the online prediction procedure.

\subsection{Overview of the ODMN Architecture}\label{sec:odmn}

\begin{figure}[htbp]
    \centering
    \includegraphics[width=0.5\textwidth]{ 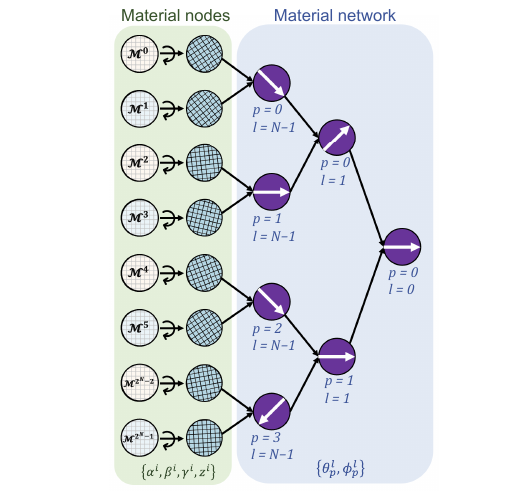}
    \caption{Schematic architecture of the ODMN. A binary-tree material network defines hierarchical stress-equilibrium interactions, and the $2^N$ material nodes represent RVE subdomains characterized by local orientations and weighting factors.}
    \label{fig:ODMN_architecture}
\end{figure}

The ODMN provides a physics-informed framework for homogenizing polycrystalline aggregates. Its formulation is rooted in micromechanical theory, particularly the Hill–Mandel condition and averaging theorems, which enable analytical homogenization within each building block. As a result, all trainable parameters in ODMN possess clear physical interpretations~\cite{Wei01}. In the context of this work, the ODMN parameter set serves as the target of the microstructure-to-parameter inference problem introduced in Section~\ref{sec2}. The overall architecture of ODMN is illustrated in Fig.~\ref{fig:ODMN_architecture}.

ODMN approximates the homogenized response of an RVE by hierarchically decomposing it into multiple subdomains, each represented by a material node. A binary-tree material network defines the interaction topology, specifying how these material nodes satisfy stress equilibrium across different levels. Each tree node is associated with parameters that determine the directions along which stress equilibrium is enforced.

The ODMN architecture is governed by a depth hyperparameter $N$ and consists of two components:
\begin{itemize}
    \item A \emph{material network} with $N$ levels, arranged as a binary tree containing $2^N - 1$ tree nodes, each defining an interaction mechanism through a stress-equilibrium direction.
    \item A set of $2^N$ \emph{material nodes}, each representing a subdomain of the RVE and characterized by local orientation and weighting parameters.
\end{itemize}

Within this hierarchy, the root interaction at the top of the binary tree corresponds to the final homogenization of the entire RVE and enforces its macroscopic equilibrium, whereas the deepest-level interactions act most locally between pairs of material nodes. Each material node represents a reduced-order constituent of the RVE rather than an individual grain; the deepest-level interactions should therefore be interpreted as interactions between representative constituents rather than between specific grains.

Each material node $\mathcal{M}^i$ is parameterized by $\{z^i, \alpha^i, \beta^i, \gamma^i\}$. The scalar $z^i$ is transformed into a non-negative weighting factor $W^i$ via the softplus function:
\begin{equation}
    W^i = \operatorname{softplus}(z^i) = \ln\left(1 + \exp\left(z^i\right)\right),
\end{equation}
which quantifies the contribution of $\mathcal{M}^i$ to the overall RVE response and implicitly governs its role in representing the crystallographic texture.

The angles $\{\alpha^i, \beta^i, \gamma^i\}$ are the Tait–Bryan angles that specify the crystallographic orientation of each material node. Each tree node is parameterized by two rotation angles, $\theta^l_p$ and $\phi^l_p$, which define the directions along which stress equilibrium is enforced at level $l$ and position $p$. The complete set of trainable parameters $\mathcal{F}$ in an ODMN of depth $N$ is given by:
\begin{equation}\label{eq:trainable_parameters}
    \begin{aligned}
        \mathcal{F} = & \left\{ z^i, \alpha^i, \beta^i, \gamma^i \mid i = 0, 1, \dots, 2^N - 1 \right\} \\
        & \cup \left\{ \theta^{l}_{p}, \phi^{l}_{p} \mid l = 0, 1, \dots, N - 1;\; p = 0, 1, \dots, 2^l - 1 \right\}
    \end{aligned}
\end{equation}

During offline training, the ODMN encodes the geometric topology of the RVE into its parameter set $\mathcal{F}$. In the classical first-order computational homogenization setting considered here, linear-elastic single-crystal stiffness matrices are sufficient to characterize microscale behavior. The resulting analytical homogenization mapping is expressed as:
\begin{equation}\label{eq:homogenization_function_at_root}
    \bar{\mathbb{C}} = \mathcal{H}_{\mathrm{ODMN}}\left(\mathbb{C}^{\mathrm{crystal}} \mid \mathcal{F}\right),
\end{equation}
\noindent where $\bar{\mathbb{C}}$ is the homogenized stiffness matrix, and $\mathcal{H}_{\mathrm{ODMN}}(\cdot)$ denotes the ODMN homogenization operator. The operator $\mathcal{H}_{\mathrm{ODMN}}$ evaluates the homogenized stiffness in two stages: a local rotation of the single-crystal stiffness at each material node, followed by a bottom-up recursive homogenization through the binary tree.

First, the single-crystal stiffness $\mathbb{C}^{\mathrm{crystal}}$ assigned to each material node $\mathcal{M}^i$ is rotated to the global frame according to that node's crystallographic orientation $(\alpha^i, \beta^i, \gamma^i)$:
\begin{equation}\label{eq:stiffness_rotation}
    \mathbb{C}^{i}_{R}
    = \mathbf{Q}_{R_1}(\alpha^i, \beta^i, \gamma^i)\,
      \mathbb{C}^{\mathrm{crystal}}\,
      \mathbf{Q}_{R_2}(\alpha^i, \beta^i, \gamma^i),
\end{equation}
\noindent where $\mathbf{Q}_{R_1}$ and $\mathbf{Q}_{R_2}$ are the rotation operators that act on the single-crystal stiffness matrix in Voigt notation~\cite{Wei01}.

The rotated stiffness matrices are then homogenized recursively, from the deepest level of the tree up to the root. At the deepest level ($l = N-1$), each interaction node merges the two rotated stiffness matrices $\mathbb{C}^{2p}_{R}$ and $\mathbb{C}^{2p+1}_{R}$ at position $p$,
\begin{equation}\label{eq:leaf_homogenization}
    \bar{\mathbb{C}}^{N-1}_{p} = \mathbb{H}_2\!\left(\mathbb{C}^{2p}_{R},\, \mathbb{C}^{2p+1}_{R},\, f^0,\, f^1,\, \vec{\mathbf{N}}^l_p\right),
\end{equation}
\noindent whereas at the higher levels ($l < N-1$) the homogenized stiffnesses of the two child nodes, $\bar{\mathbb{C}}^{l+1}_{2p}$ and $\bar{\mathbb{C}}^{l+1}_{2p+1}$, are recursively combined,
\begin{equation}\label{eq:recursive_homogenization}
    \bar{\mathbb{C}}^{l}_{p} = \mathbb{H}_2\!\left(\bar{\mathbb{C}}^{l+1}_{2p},\, \bar{\mathbb{C}}^{l+1}_{2p+1},\, f^0,\, f^1,\, \vec{\mathbf{N}}^l_p\right),
\end{equation}
\noindent so that the stiffness returned at the root, $\bar{\mathbb{C}} \equiv \bar{\mathbb{C}}^{0}_{0}$, is the homogenized response of the entire RVE.

Each two-branch operation is carried out by the binary homogenization operator $\mathbb{H}_2$, which enforces stress equilibrium and kinematic compatibility across the interface between the two branches:
\begin{equation}\label{eq:H2_function}
\begin{aligned}
\bar{\mathbb{C}} = \mathbb{H}_2(\mathbb{C}^0, \mathbb{C}^1, f^0, f^1, \vec{\mathbf{N}})
= {} & f^0 \mathbb{C}^0 + f^1 \mathbb{C}^1 \\
& - f^0 f^1 (\mathbb{C}^0 - \mathbb{C}^1)\,
\mathbf{Q}(\vec{\mathbf{N}}, f^0, f^1)\,(\mathbb{C}^0 - \mathbb{C}^1),
\end{aligned}
\end{equation}
\noindent where $\mathbb{C}^0$ and $\mathbb{C}^1$ are the stiffness matrices of the two branches and $f^0, f^1$ are their volume fractions. The interaction matrix $\mathbf{Q}$ depends on the interface normal $\vec{\mathbf{N}}$ and the volume fractions through
\begin{equation}\label{eq:Q_matrix}
    \mathbf{Q}
    =
    \mathbf{H}\mathbf{S}^{-1}\mathbf{H}^{\top},
    \qquad
    \mathbf{S}
    =
    \mathbf{H}^{\top}
    \left(
        f^{1}\mathbb{C}^{0}
        +
        f^{0}\mathbb{C}^{1}
    \right)
    \mathbf{H}.
\end{equation}
\noindent in which the matrix $\mathbf{H}(\vec{\mathbf{N}})$ maps the interface normal $\vec{\mathbf{N}} = (N_1, N_2, N_3)$ into Voigt notation,
\begin{equation}\label{eq:H_matrix}
    \mathbf{H}(\vec{\mathbf{N}}) =
    \begin{bmatrix}
N_1 & 0   & 0 \\
0   & N_2 & 0 \\
0   & 0   & N_3 \\
0   & N_3 & N_2 \\
N_3 & 0   & N_1 \\
N_2 & N_1 & 0
    \end{bmatrix}.
\end{equation}
The interface normal at each interaction node is parameterized by the two tree-node angles $(\theta^l_p, \phi^l_p)$,
\begin{equation}\label{eq:normal_from_angles}
    \vec{\mathbf{N}}^l_p = \vec{\mathbf{N}}\!\left(\phi^l_p, \theta^l_p\right),
\end{equation}
\noindent which sets the direction along which stress equilibrium is enforced at that node. Finally, the branch volume fractions are obtained by aggregating the weighting factors of the material nodes belonging to each branch,
\begin{equation}\label{eq:branch_fractions}
    f^0 = \frac{\sum_{i \in \mathcal{K}^0} W^i}{\sum_{i \in \mathcal{K}^0 \cup \mathcal{K}^1} W^i},
    \qquad
    f^1 = \frac{\sum_{i \in \mathcal{K}^1} W^i}{\sum_{i \in \mathcal{K}^0 \cup \mathcal{K}^1} W^i},
\end{equation}
\noindent where $\mathcal{K}^0$ and $\mathcal{K}^1$ denote the sets of material-node indices belonging to the two branches.

\subsection{TACS Initialization Strategy}\label{sec:tacs}

Within the proposed TACS--GNN--ODMN framework, the TACS method~\cite{wanni2024machine} is employed to represent the crystallographic texture in the microstructure-to-parameter inference problem. Specifically, TACS constructs a reduced yet statistically consistent set of representative orientations that retains the essential features of the original ODF of a polycrystalline aggregate.

The procedure starts with K-means clustering in orientation space. The number of clusters is selected using a within-cluster sum of squares (WCSS) criterion to balance resolution and computational efficiency. A density-aware sampling scheme is then applied within each cluster to represent both dense and sparse regions. The reduced orientation set is further refined by comparing the reconstructed ODF histogram with the original distribution. This refinement continues until the relative deviation falls below a prescribed tolerance. The final set of Tait--Bryan angles provides a compact and accurate representation of the crystallographic texture.

For an ODMN with hyperparameter $N$, the network contains $2^N$ material nodes. Accordingly, TACS samples exactly $2^N$ orientations from the reconstructed ODF. These orientations are directly assigned to the crystallographic parameters $\{\alpha^i, \beta^i, \gamma^i \mid i = 0, 1, \dots, 2^N - 1\}$ in $\mathcal{F}$, and the corresponding weighting factors are uniformly initialized as $z^i = 1$. This initialization ensures a physically consistent representation of the initial crystallographic texture. It also establishes a stable basis for subsequent parameter inference and improves the prediction of texture evolution during online simulation.

Through this procedure, the full-resolution ODF of the polycrystalline aggregate is compressed into a compact set of $2^N$ representative orientations, whose accuracy in reproducing the original ODF is quantified for the test microstructures in Section~\ref{sec:offline_results}.

\subsection{Microstructure Graph Construction}\label{sec:graph}

To enable microstructure-to-parameter inference, the polycrystalline RVE is represented as a grain-level graph \(G = (V, E)\), where nodes encode grain attributes and edges capture intergranular interactions. Each node corresponds to a grain, and each edge represents a grain-boundary contact, with connectivity defined by the shared-boundary topology.

Each grain’s crystallographic orientation is mapped to the face-centered cubic (FCC) fundamental zone and represented as a quaternion \((q_0, q_1, q_2, q_3)\), avoiding the singularities associated with Euler angles. In addition to orientation, geometric descriptors are incorporated to characterize grain morphology, including centroid coordinates \((x_\mathrm{c}, y_\mathrm{c}, z_\mathrm{c})\), a periodicity flag \(P_\mathrm{grain}\), and the components of the second moment tensor \((I_{xx}, I_{yy}, I_{zz}, I_{xy}, I_{yz}, I_{zx})\).

To establish a consistent correspondence between grains and ODMN material nodes, we introduce a permutation-invariant orientation index. Since the TACS-sampled orientations \(\{\mathbf{q}_i\}_{i=0}^{2^N-1}\) are unordered and may vary across sampling runs, each grain is matched to the closest orientation in the reduced TACS set. The orientation index is defined as
\begin{equation}
    \mathrm{idx}_\mathrm{ori} = \argmin_{i} \; d_\mathrm{geo}\!\left(\mathbf{q}_\mathrm{grain}, \mathbf{q}_i\right),
\end{equation}
\noindent where the geodesic distance is
\begin{equation}
    d_\mathrm{geo}(\mathbf{q}_a, \mathbf{q}_b) = 2 \arccos \left( \left| \langle \mathbf{q}_a, \mathbf{q}_b \rangle \right| \right).
\end{equation}

This nearest-orientation assignment establishes a deterministic, physically consistent alignment between grains and ODMN material nodes, eliminating ambiguities due to permutation variability in TACS sampling.
The complete set of node features is summarized in Table~\ref{tab:node_features}.

\begin{table}[htbp]
\centering
\caption{Grain-level node features for microstructure graph representation.}
\label{tab:node_features}
\small
\setlength{\tabcolsep}{5pt}
\begin{tabularx}{\linewidth}{l p{2.9cm} X}
\toprule
Feature & Symbol / Unit & Description \\
\midrule
Quaternion components 
& \((q_0, q_1, q_2, q_3)\) 
& Crystallographic orientation in the FCC fundamental zone, represented by a unit quaternion. \\

Normalized volume 
& \(V_\mathrm{grain}/V_\mathrm{RVE}\) 
& Grain volume fraction relative to the RVE. \\

Periodicity flag 
& \(P_\mathrm{grain}\) 
& Boolean indicator equal to 1 if the grain spans a periodic boundary of the RVE, and 0 otherwise. \\

Centroid coordinates 
& \((x_\mathrm{c}, y_\mathrm{c}, z_\mathrm{c})\) 
& Grain centroid position in Cartesian coordinates. \\

Second moment components 
& \(I_{xx}, \ldots, I_{zx}\) 
& Components of the grain inertia tensor characterizing morphological anisotropy. \\

Orientation index 
& \(\mathrm{idx}_\mathrm{ori}\) 
& Index \(i\) of \(\mathbf{q}_i\) in \(\{\mathbf{q}_i\}\) with the minimum geodesic distance to the grain orientation \(\mathbf{q}_\mathrm{grain}\). \\
\bottomrule
\end{tabularx}
\end{table}

\subsection{GNN Architecture}\label{sec:gnn}

The GNN module serves as the inference operator that maps the grain-level microstructure graph to the micromechanical equilibrium parameters in \(\mathcal{F}\). Specifically, it predicts the parameter set 
\(\{\theta^{l}_{p}, \phi^{l}_{p} \mid l = 0, 1, \dots, N - 1;\; p = 0, 1, \dots, 2^l - 1\}\), 
which contains \(2^{N+1} - 2\) angular components corresponding to the internal nodes of the ODMN hierarchy. Graph neural networks are well-suited to this inference task, as they operate directly on the irregular connectivity of grain interaction graphs and have proven effective for mechanics and structural analysis problems~\cite{chou2024structgnn}.

The architecture employs two stacked GATv2Conv layers~\cite{brody2022how} to aggregate information from neighboring grains, followed by rectified linear unit (ReLU) activations. GATv2Conv implements the second version of the graph attention network (GAT). Compared with the original GAT, it applies a shared linear transformation \(\Theta_s\) to both source and target node features prior to attention computation, allowing each node to adaptively weight its neighbors and improving expressiveness in graphs with complex connectivity.

The attention coefficient between nodes \(i\) and \(j\) is defined as
\begin{equation}
    \alpha_{ij} = 
\frac{\exp\!\left(a^{\top} \,\mathrm{LeakyReLU}\!\left(\Theta_s x_i + \Theta_s x_j\right)\right)}
{\sum_{k \in \mathcal{N}(i) \cup \{i\}} 
 \exp\!\left(a^{\top} \,\mathrm{LeakyReLU}\!\left(\Theta_s x_i + \Theta_s x_k\right)\right)},
\end{equation}
and the updated node representation is
\begin{equation}
    x'_i = \sum_{j \in \mathcal{N}(i) \cup \{i\}} \alpha_{ij} \, \Theta_s x_j .
\end{equation}
\noindent Here, \(x_i\) denotes the feature vector of node \(i\), \(a\) is a learnable attention vector, and \(\mathcal{N}(i)\) is the set of neighboring grains directly connected to grain \(i\) by graph edges, corresponding to grains that share a grain-boundary contact with grain \(i\). The union \(\mathcal{N}(i)\cup\{i\}\) indicates that self-information is included in the attention aggregation. The activation function \(\mathrm{LeakyReLU}(\cdot)\) is applied element-wise and is defined as
\begin{equation}
    \mathrm{LeakyReLU}(z) =
    \begin{cases}
        z, & z \geq 0, \\
        \eta z, & z < 0,
    \end{cases}
\end{equation}
\noindent where \(\eta=0.2\) is the negative-slope coefficient used in this work.

After graph convolution, a global mean pooling operation aggregates node-level embeddings into a graph-level representation. This representation is passed through a regression head consisting of a fully connected layer and a Softplus activation, producing the \(2^{N+1} - 2\) nonnegative angular parameters of the material network. The overall architecture is summarized in Table~\ref{tab:gnn_architecture}.

\begin{table}[htbp]
\centering
\caption{Architecture of the GNN module for predicting ODMN micromechanical equilibrium parameters. 
\(B\) denotes the number of graphs in a batch, and \(n_{\mathrm{nodes}}\) the number of grains per graph.}
\label{tab:gnn_architecture}
\small
\setlength{\tabcolsep}{4pt}
\begin{tabularx}{\linewidth}{p{2.6cm} X p{3.2cm}}
\hline
Stage & Layer / Operation & Output shape \\
\hline
Input stage & Grain-level feature matrix & \((B, n_{\mathrm{nodes}}, 16)\) \\
\hline
\multirow{4}{*}{\parbox{2.6cm}{\centering Graph convolution stage}}
& GATv2Conv \(16 \rightarrow 64,\ \mathrm{heads}=1\) & \((B, n_{\mathrm{nodes}}, 64)\) \\
& ReLU activation & \((B, n_{\mathrm{nodes}}, 64)\) \\
& GATv2Conv \(64 \rightarrow 32,\ \mathrm{heads}=1\) & \((B, n_{\mathrm{nodes}}, 32)\) \\
& ReLU activation & \((B, n_{\mathrm{nodes}}, 32)\) \\
\hline
Pooling stage & Global mean pooling & \((B, 32)\) \\
\hline
\multirow{2}{*}{\parbox{2.6cm}{\centering Regression head}}
& Fully connected \(32 \rightarrow (2^{N+1} - 2)\) & \((B, 2^{N+1} - 2)\) \\
& Softplus activation & \((B, 2^{N+1} - 2)\) \\
\hline
\end{tabularx}
\end{table}

\begin{figure}[htbp]
    \centering
    \includegraphics[width=1\textwidth]{ 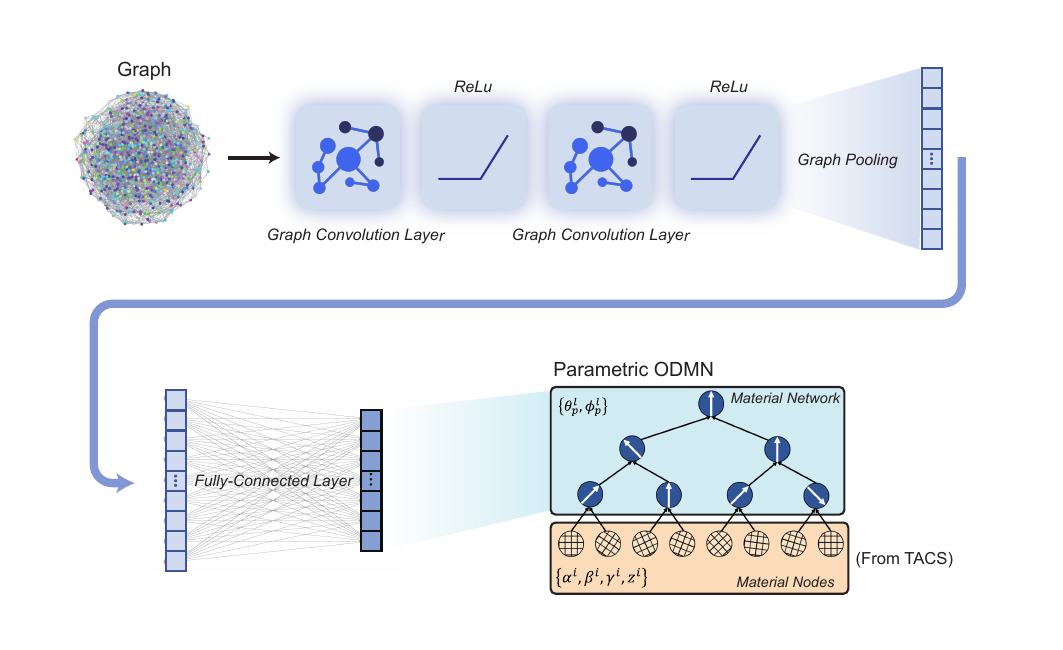}
    \caption{Architecture of the GNN module for predicting ODMN micromechanical equilibrium parameters. 
A grain-level microstructure graph is processed by two GATv2Conv layers with ReLU activations, followed by global mean pooling and a fully connected layer with Softplus activation to output the angular parameters \(\{\theta^{l}_{p}, \phi^{l}_{p}\}\) of the material network.}
    \label{fig:model_arch}
\end{figure}

\subsection{Dataset Preparation}\label{sec:dataset}

To enable generalization across diverse microstructures, the training dataset is constructed to cover a broad range of crystallographic textures and corresponding mechanical responses. Each polycrystalline RVE is characterized by its ODF, computed using DREAM.3D (v6.5.171)~\cite{groeber2014dream}. The ODF is discretized into 1,000 orientations in the FCC fundamental zone. Each orientation is specified by a set of Euler angles, a weight in multiples of random density (MRD), and a spread parameter. The spread parameter controls the blurring of the selected orientation in Rodrigues space and defines the number of orientation bins over which the MRD value decreases quadratically to zero.

Based on this representation, four representative texture classes are constructed following~\cite{DAI2021117006}, spanning both strong and weak texture regimes. For each class, 360 ODF realizations are generated, resulting in a total of 1,440 RVEs. Each RVE contains approximately 809 grains and is discretized on a \(45 \times 45 \times 45\) voxel grid.

To account for variability in mechanical response, 500 sets of elastic constants $\{C_{11}, C_{12}, C_{44}\}$ are sampled for each RVE to construct cubic single-crystal stiffness matrices $\mathbb{C}^{\mathrm{crystal}}$ in Voigt notation:
\begin{equation}\label{eq:crystal_stiffness_matrix}
\mathbb{C}^{\mathrm{crystal}} =
\begin{bmatrix}
C_{11} & C_{12} & C_{12} & 0 & 0 & 0 \\
C_{12} & C_{11} & C_{12} & 0 & 0 & 0 \\
C_{12} & C_{12} & C_{11} & 0 & 0 & 0 \\
0 & 0 & 0 & C_{44} & 0 & 0 \\
0 & 0 & 0 & 0 & C_{44} & 0 \\
0 & 0 & 0 & 0 & 0 & C_{44}
\end{bmatrix}.
\end{equation}

The sampling of $\{C_{11}, C_{12}, C_{44}\}$ follows the protocol in the ODMN study~\cite{Wei01}. For each sample, full-field DNS is performed using the FFT-based DAMASK solver~\cite{roters2019damask} to obtain the corresponding homogenized stiffness matrix, $\mathbb{C}^{\mathrm{DNS}}$. All simulations were carried out with DAMASK~3.0.0-alpha4 using the default solver convergence tolerances,
which are listed in Table~\ref{tab:damask_tol}.

\begin{table}[htbp]
\centering
\caption{Default convergence tolerances of the FFT-based mechanical solver in
DAMASK~3.0.0-alpha4 used for generating the training dataset.}
\label{tab:damask_tol}
\begin{tabular}{l p{5.4cm} c l}
\hline
Parameter & Meaning & Value & Unit \\
\hline
\texttt{eps\_div\_atol}    & Abs.\ tolerance for equilibrium residual & $1.0\times10^{-4}$ & m$^{-1}$ \\
\texttt{eps\_div\_rtol}    & Rel.\ tolerance for equilibrium residual & $5.0\times10^{-4}$ & -- \\
\texttt{eps\_stress\_atol} & Abs.\ tolerance for stress boundary-condition residual    & $1.0\times10^{3}$  & Pa \\
\texttt{eps\_stress\_rtol} & Rel.\ tolerance for stress boundary-condition residual    & $1.0\times10^{-3}$ & -- \\
\texttt{itmin}             & Minimum iterations per increment          & $1$                & -- \\
\texttt{itmax}             & Maximum iterations per increment          & $250$              & -- \\
\hline
\end{tabular}
\end{table}

The final dataset consists of 1,280 RVEs for training and 160 RVEs for validation, with all texture classes equally represented. Each RVE is associated with 500 pairs of stiffness data, $\{  \mathbb{C}^{\mathrm{crystal}}, \mathbb{C}^{\mathrm{DNS}} \}$. This dataset captures both microstructural variability and the corresponding mechanical response, enabling the model to learn a consistent microstructure-to-parameter mapping. In addition, four unseen RVEs are used to evaluate the generalization performance of the proposed framework.

\subsection{Offline Training Procedure}\label{sec:offline}

The TACS--GNN--ODMN framework is trained end-to-end to learn a mapping from microstructures to ODMN parameters, which subsequently define the corresponding homogenized response. For each training sample, the RVE is first processed by the TACS module, which extracts $2^N$ representative orientations from the ODF to initialize the ODMN texture parameters $\{\alpha^i, \beta^i, \gamma^i\}$ with uniform weights $z^i = 1$. Based on these representative orientations, a grain-level microstructure graph $G$ is constructed and passed to the GNN, which predicts the micromechanical equilibrium parameters $\{\theta^{l}_{p}, \phi^{l}_{p}\} \in \mathcal{F}$.

Together, these parameters define a fully parameterized ODMN for the given microstructure and enable the analytical homogenization mapping
\begin{equation}
    \bar{\mathbb{C}} = \mathcal{H}_{\mathrm{ODMN}}\left(
        \mathbb{C}^{\mathrm{crystal}} \;\middle|\;
        \{z^i, \alpha^i, \beta^i, \gamma^i\},\;
        \{\theta^{l}_{p}, \phi^{l}_{p}\} = f_{\mathrm{GNN}}(G)
    \right),
\end{equation}
\noindent where $\mathbb{C}^{\mathrm{crystal}}$ denotes the single-crystal stiffness matrix and $\bar{\mathbb{C}}$ is the predicted homogenized stiffness matrix.

The predicted response $\bar{\mathbb{C}}^{\mathrm{ODMN}}$ is compared with the reference response $\bar{\mathbb{C}}^{\mathrm{DNS}}$, obtained from FFT-based DNS in DAMASK~\cite{roters2019damask}, using a relative Frobenius norm loss:
\begin{equation}
    \mathcal{L} =
    \frac{1}{N_{\mathrm{batch}}}
    \sum_{i=1}^{N_{\mathrm{batch}}}
    \frac{\left \| \bar{\mathbb{C}}^{\mathrm{DNS}}_i - \bar{\mathbb{C}}^{\mathrm{ODMN}}_i \right \|^2}
    {\left \| \bar{\mathbb{C}}^{\mathrm{DNS}}_i \right \|^2},
\end{equation}
\noindent where $N_{\mathrm{batch}}$ is the number of samples in a mini-batch, $\bar{\mathbb{C}}^{\mathrm{DNS}}_i$ and $\bar{\mathbb{C}}^{\mathrm{ODMN}}_i$ are the reference and predicted homogenized stiffness matrices of the $i$-th sample, and $\|\cdot\|$ denotes the Frobenius norm.

The model parameters are optimized using AdamW, with early stopping based on the validation loss.

\subsection{Online Prediction Procedure}\label{sec:online}

During online prediction, an unseen RVE is processed through the same two-branch decomposition. The TACS module initializes the texture parameters, while the GNN infers the equilibrium parameters from the corresponding microstructure graph, thereby defining a microstructure-specific ODMN. The resulting ODMN is then used to predict the homogenized response and texture evolution under prescribed loading paths, following the online prediction algorithm of the ODMN~\cite{Wei01}. Within the ODMN, nonlinear homogenization is carried out by a localization--constitutive-update--homogenization cycle embedded in a Newton--Raphson scheme, which efficiently approximates the full-field nonlinear response while preserving micromechanical consistency.

At each macroscopic material point, the ODMN receives the deformation gradient $\bar{\mathbf{F}}$ imposed by the macro-scale analysis. The macroscopic deformation is first localized to the material nodes, the crystal-plasticity constitutive response is then updated locally at each node, and the resulting nodal stresses are homogenized to obtain the macroscopic first Piola--Kirchhoff stress $\bar{\mathbf{P}}$. The interface fluctuation variables are determined by solving the ODMN equilibrium problem formulated to satisfy Hill--Mandel consistency. Upon convergence, the homogenized stress and the consistent tangent $\partial \bar{\mathbf{P}}/\partial \bar{\mathbf{F}}$ are supplied to the outer macro-scale Newton solver.

The cycle begins with the \emph{downscaling}, or localization, step, in which the macroscopic deformation gradient is distributed to the $2^N$ material nodes through the interaction-based mapping \begin{equation}\label{eq:interaction_mapping}
   \mathbf{F}^i = \bar{\mathbf{F}} + \sum_{l=0}^{N-1}
\sum_{p=0}^{2^{l}-1} \alpha^{i}_{l,p}\, \mathbf{a}^{l}_{p} \otimes \mathbf{N}^{l}_{p},
   \qquad i = 0, 1, \dots, 2^{N}-1 .
\end{equation}
\noindent Here, the interaction coefficients $\alpha^{i}_{l,p}$ are determined by the material-node weighting factors $\{W^i\}$, i.e., the volume fractions, together with the binary-tree topology. The vector $\mathbf{N}^{l}_{p}$ denotes the interface normal associated with tree node $(l,p)$ and is prescribed by the trained angles $(\theta^l_p,\phi^l_p)$. In contrast, the interface fluctuation vectors $\mathbf{a}^{l}_{p}$ are not learned offline. They are the unknowns of the online localization problem and are solved iteratively by enforcing the ODMN equilibrium conditions required for Hill--Mandel consistency.

For a given set of interface fluctuation vectors $\{\mathbf{a}^{l}_{p}\}$, Eq.~\eqref{eq:interaction_mapping} provides the local deformation gradient $\mathbf{F}^i$ at each material node. The corresponding local stress is then evaluated using a finite-strain phenomenological crystal-plasticity model. The deformation gradient at node $i$ is multiplicatively decomposed into elastic and plastic parts,
\begin{equation}\label{eq:multiplicative_decomposition_online}
    \mathbf{F}^i = \mathbf{F}^i_e \mathbf{F}^i_p ,
\end{equation}
\noindent where $\mathbf{F}^i_p$ maps the reference configuration to an intermediate, plastically deformed configuration, and $\mathbf{F}^i_e$ describes the subsequent elastic stretch and lattice rotation. The elastic response is governed by the generalized Hooke's law,
\begin{equation}\label{eq:hooke_online}
    \mathbf{S}^i = \mathbb{C} : \mathbf{E}^i,
    \qquad
    \mathbf{E}^i = \frac{1}{2} \left[ (\mathbf{F}^i_e)^{\top}\mathbf{F}^i_e - \mathbf{I} \right],
\end{equation}
\noindent where $\mathbf{S}^i$ is the second Piola--Kirchhoff stress, $\mathbf{E}^i$ is the elastic Green--Lagrange strain, and $\mathbb{C}$ is the single-crystal elastic stiffness. The plastic deformation evolves according to
\begin{equation}\label{eq:flow_rule_online}
    \dot{\mathbf{F}}^i_p = \mathbf{L}^i_p \mathbf{F}^i_p ,
\end{equation}
\noindent with the plastic velocity gradient expressed as the sum of shear-rate contributions over all slip systems,
\begin{equation}\label{eq:plastic_velocity_online}
    \mathbf{L}^i_p = \sum_{\alpha} \dot{\gamma}^{i,\alpha} \left( \mathbf{s}^{\alpha}_s \otimes \mathbf{n}^{\alpha}_s \right).
\end{equation}
\noindent Here, the shear rate $\dot{\gamma}^{i,\alpha}$ follows a viscoplastic power law of the resolved shear stress $\tau^{i,\alpha}$ and the slip resistance $\xi^{i,\alpha}$. The resolved shear stress is obtained from Schmid's law applied to the Mandel stress, while the slip resistance evolves according to a saturation-type hardening law. The explicit slip and hardening relations are provided in~\ref{appendixA}.

We collect the interface fluctuation vectors of all internal nodes of the binary tree into the set
\begin{equation}
\label{eq:all_fluctuation_vectors}
\mathcal{A}
\coloneqq
\left\{
\mathbf{a}^{l}_{q}
\;\middle|\;
l = 0,\ldots,N-1,\;
q = 0,\ldots,2^{l}-1
\right\}.
\end{equation}
The vectors in $\mathcal{A}$ are determined by enforcing the Hill--Mandel consistency condition. In the ODMN, the rank-one localization ansatz reduces this condition to a traction-balance equation for each interaction mechanism. For the interface associated with tree node $(l,p)$, the balance reads 
\begin{equation}
\label{eq:interaction_equilibrium_mech}
\mathbf{r}^{\,l}_{p}
\coloneqq
\sum_{i=0}^{2^{N}-1}
W^{i}\,
\alpha^{i}_{l,p}\,
\mathbf{P}^{i}
\!\left(
\mathbf{F}^{i}(\bar{\mathbf{F}},\mathcal{A})
\right)
\cdot
\mathbf{N}^{\,l}_{p}
=
\mathbf{0},
\qquad
l = 0,\ldots,N-1,\quad
p = 0,\ldots,2^{l}-1 .
\end{equation}
The coupled nonlinear residual system $\{\mathbf{r}^{\,l}_{p}=\mathbf{0}\}$ is solved for the interface fluctuation vectors $\mathcal{A}$ using Newton--Raphson iteration. Upon convergence of this inner equilibrium solve, the homogenized first Piola--Kirchhoff stress $\bar{\mathbf{P}}$ and the consistent tangent $\partial\bar{\mathbf{P}}/\partial\bar{\mathbf{F}}$, required by the outer macroscale Newton solver, are obtained analytically following the original ODMN formulation~\cite{Wei01}.

Throughout the incremental analysis, a clear separation is maintained between the calibrated network parameters and the quantities updated at each load step. The GNN-predicted interface angles $\{\theta^l_p,\phi^l_p\}$ and the material-node weighting factors $\{W^i\}$ are determined once during offline training and remain fixed, as they encode the homogenization topology and interface geometry of the RVE. At every increment, only the interface fluctuation vectors $\mathcal{A}$ and the crystal-plasticity internal variables at each material node are recomputed. The crystallographic orientations obtained offline serve solely as the initial lattice orientations; once deformation begins, the lattice rotation of each node evolves through the constitutive update and is recovered from the polar decomposition $\mathbf{F}^i_e = \mathbf{R}^i_t\,\mathbf{U}^i_t$. The homogenized stiffness, which depends on these orientations through Eq.~\eqref{eq:stiffness_rotation}, therefore evolves with deformation, whereas the underlying network parameters do not.

\section{Results and Discussion}\label{sec3}

\subsection{Offline Training Results}\label{sec:offline_results}

\begin{figure}[htbp]
    \centering
    \includegraphics[width=1.0\linewidth]{ 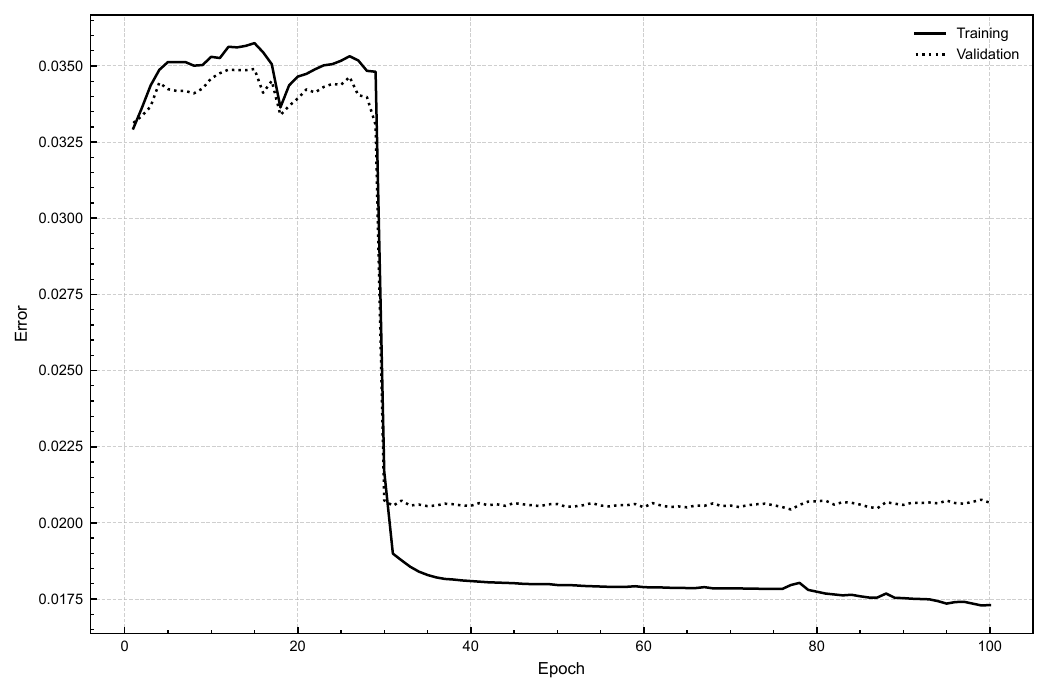}
    \caption{Training and validation error curves of the TACS--GNN--ODMN framework with \(N = 7\).}
    \label{fig:TrainingCurve}
\end{figure}

The offline training results are first examined in terms of convergence behavior and generalization to unseen microstructures. The framework is trained with an ODMN depth of \(N = 7\). Fig.~\ref{fig:TrainingCurve} shows the evolution of training and validation errors over 100 epochs. Both errors decrease rapidly during the initial stage and gradually stabilize, indicating effective learning of the microstructure-to-parameter mapping. The validation error reaches its minimum value of 0.0205 at epoch 35, and the corresponding model is used for all subsequent evaluations.

\begin{figure}[htbp]
    \centering
    \includegraphics[width=1.0\linewidth]{ 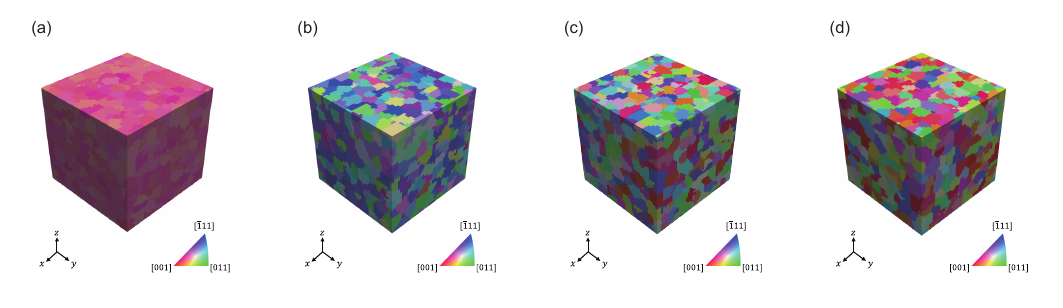}
    \caption{Unseen RVEs used for testing: (a) S1, (b) S2, (c) W1, (d) W2.}
    \label{fig:TestingRVEs}
\end{figure}

To assess generalization beyond the training set, four unseen polycrystalline RVEs are generated, denoted as S1 (strong-textured-1), S2 (strong-textured-2), W1 (weak-textured-1), and W2 (weak-textured-2) (Fig.~\ref{fig:TestingRVEs}). These RVEs span both strongly and weakly textured regimes, providing a stringent test of the proposed framework under a wide range of texture distributions.

\begin{figure}[htbp]
    \centering
    \includegraphics[width=1\linewidth]{ 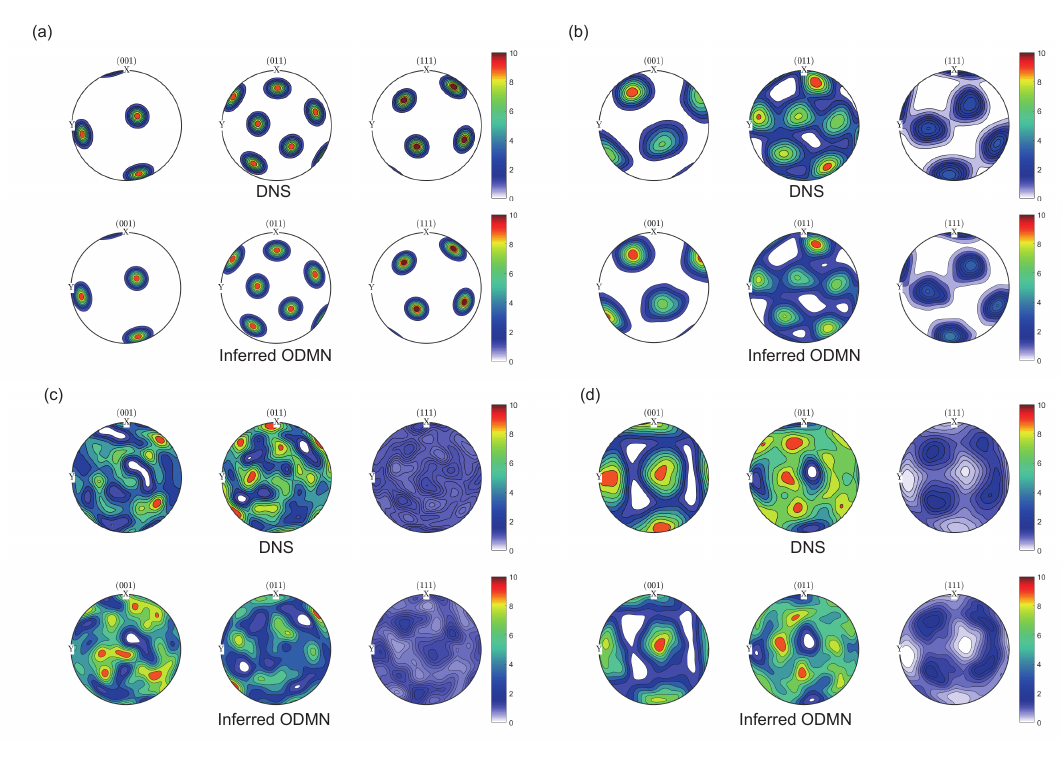}
    \caption{Pole figures of reconstructed ODFs compared with DNS results for the four testing RVEs: (a) S1, (b) S2, (c) W1, and (d) W2.}
    \label{fig:initial_texture}
\end{figure}

Each testing RVE is processed through the TACS--GNN--ODMN framework to construct the corresponding ODMN parameters. The inferred crystallographic parameters \(\{\alpha^i, \beta^i, \gamma^i \mid i = 0, 1, \dots, 2^N - 1\}\) are used to reconstruct the ODFs, which are visualized as pole figures in Fig.~\ref{fig:initial_texture}. The reconstructed ODFs exhibit excellent agreement with DNS results, accurately capturing both dominant orientation clusters and their dispersion.

To quantify the reconstruction accuracy, the normalized texture index of the difference ODF (DODF), \(\hat{T}^d\), is evaluated:
\begin{equation}
    \hat{T}^d = \frac{\int [f_{\mathrm{ODMN}}(g) - f_{\mathrm{DNS}}(g)]^2 \, dg}
                     {\int [f_{\mathrm{DNS}}(g)]^2 \, dg},
\end{equation}
\noindent where \(f_{\mathrm{ODMN}}(g)\) and \(f_{\mathrm{DNS}}(g)\) denote the ODF reconstructed from the ODMN and the reference ODF obtained from DNS, respectively.

The resulting \(\hat{T}^d\) values are summarized in Table~\ref{tab:texture_index_initial}. All cases exhibit low \(\hat{T}^d\), confirming that the proposed framework accurately reconstructs the underlying texture without retraining. Relatively larger errors are observed for weakly textured cases, which exhibit broader orientation distributions that are inherently more difficult to approximate using a compact set of representative orientations. This behavior is consistent with observations in the original ODMN formulation~\cite{Wei01}. Nevertheless, even the most challenging case satisfies \(\hat{T}^d < 0.11\), demonstrating the robustness and generalization capability of the proposed framework across a wide range of texture intensities.

\begin{table}[htbp]
\centering
\caption{Normalized texture index of the DODF $\hat{T}^d$ for the four testing RVEs.}
\begin{tabular}{lc}
\hline
Microstructure & $\hat{T}^d$ \\
\hline
S1 & 0.0001 \\
S2 & 0.0436 \\
W1 & 0.1036 \\
W2 & 0.0847 \\
\hline
\end{tabular}
\label{tab:texture_index_initial}
\end{table}

\subsection{Robustness to Random Initialization and Training Stability}\label{sec:seed_study}

\begin{figure}[htbp]
    \centering
    \includegraphics[width=1.0\linewidth]{ 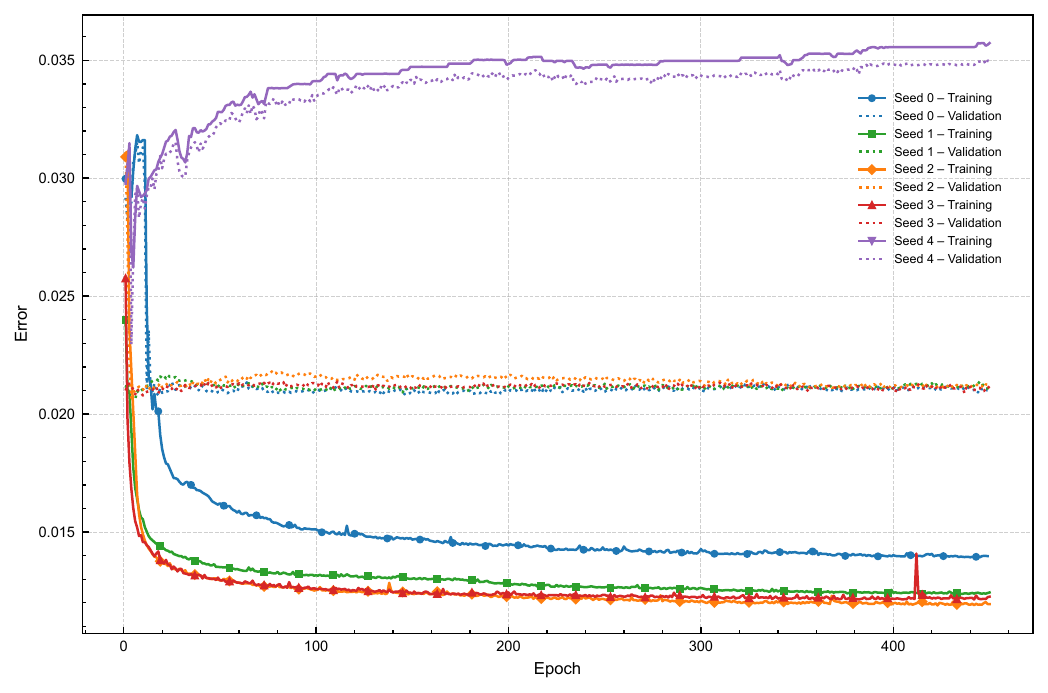}
    \caption{Training (solid lines) and validation (dotted lines) error histories of the TACS--GNN--ODMN framework ($N = 7$) for five independent random initializations (seeds $0$--$4$).}
    \label{fig:seed_training}
\end{figure}

To assess the sensitivity of the proposed framework to the random initialization of the network parameters and to verify the stability of the training loss over an extended training horizon, the offline training procedure described in Section~\ref{sec:offline} was repeated using five independent random seeds (seeds $0$--$4$). All runs used the same network depth ($N = 7$) and the same dataset. For this additional study, a larger batch size of $4096$ was adopted on an NVIDIA H200 graphics processing unit (GPU) to improve throughput; this choice affects only the optimization efficiency and not the converged mapping. Each run was trained for 500 epochs, which is five times the number of epochs used for the model reported in the main results. The resulting training and validation error histories are shown in Fig.~\ref{fig:seed_training}.

As shown in Fig.~\ref{fig:seed_training}, four of the five runs, corresponding to seeds $0$--$3$, converge rapidly and consistently. In these cases, the training error decreases below $0.015$, while the validation error stabilizes at approximately $0.021$ within the first few tens of epochs. After convergence, both the training and validation curves remain essentially flat throughout the entire 500-epoch training horizon, with no indication of overfitting or numerical divergence. The converged validation errors are also consistent with the minimum validation error of $0.0205$ reported earlier for the model used in the main results. The close agreement among these four independent runs indicates that the learned microstructure-to-parameter mapping is largely insensitive to the random initialization of the network parameters.

The remaining run, corresponding to seed $4$, does not converge to the same solution basin. Its error remains substantially higher than those of the converged runs and gradually increases during training, suggesting an unfavorable initialization from which the optimizer could not recover. In practice, such non-converging initializations are readily identified within the first few tens of epochs and can be remedied by restarting the training with a different random seed. Overall, the five-seed study shows that the proposed framework exhibits stable convergence behavior for most random initializations, and that the converged solutions remain stable over a training horizon substantially longer than that used for the reported model.

\subsection{Online Prediction Results}

The online prediction results evaluate the ability of the TACS--GNN--ODMN framework to generalize the learned microstructure-to-parameter mapping to unseen RVEs under nonlinear deformation. The four inferred ODMNs are examined using a constitutive framework that combines a phenomenological crystal plasticity model with generalized Hooke’s law to describe the elastic–plastic response. The constitutive parameters assigned to each material node are adopted from \citet{barrett2019deep} and the DAMASK documentation~\cite{damask_documentation}. The corresponding elastic and plastic parameters are summarized in Table~\ref{tab:crystal_plasticity_parameters}. The detailed crystal plasticity formulation is given in~\ref{appendixA}, and the elastic constitutive relations are provided in~\ref{appendixB}.

\begin{table}[htbp]
\centering
\caption{Elastic and plastic material parameters for AA6022-T4 aluminum alloy.}
\label{tab:crystal_plasticity_parameters}
\renewcommand{\arraystretch}{1.2}
\setlength{\tabcolsep}{6pt}
\begin{tabular}{cccccccc}
\hline
\textbf{$N_s$} & \textbf{$h_0^{\text{sl-sl}}$}(GPa) & \textbf{$\xi^{\alpha}_{\infty}$}(MPa) & \textbf{$\xi^{0}$}(MPa) & \textbf{$n$} & \textbf{$a$} & \textbf{$\dot{\gamma}_0$}$(\text{s}^{-1})$ & \textbf{$h_{\text{int}}^{\alpha}$} \\
\hline
12 & 1.02 & 266 & 76 & 20 & 3.7 & 0.001 & 0 \\
\hline
\end{tabular}

\vspace{0.5cm}

\begin{tabular}{cccc}
\hline
{$C_{11}$ (GPa)} & {$C_{12}$ (GPa)} & {$C_{44}$ (GPa)} & {$h^{\text{sl-sl}}$} \\
\hline
107.3 & 60.8 & 28.3 & [1, 1, 5.123, 0.574, 1.123, 1.123, 1] \\
\hline
\end{tabular}
\end{table}

Two loading conditions are considered: uniaxial cyclic loading along the $x$-direction with a strain rate of $\dot{F}_{11}=1$, and simple shear deformation with a strain rate of $\dot{F}_{21}=1$. For both cases, ODMN predictions are directly compared with DNS results.

\subsubsection{Cyclic loading: mechanical prediction}

\begin{figure}[htbp]
    \centering
    \includegraphics[width=1\linewidth]{ 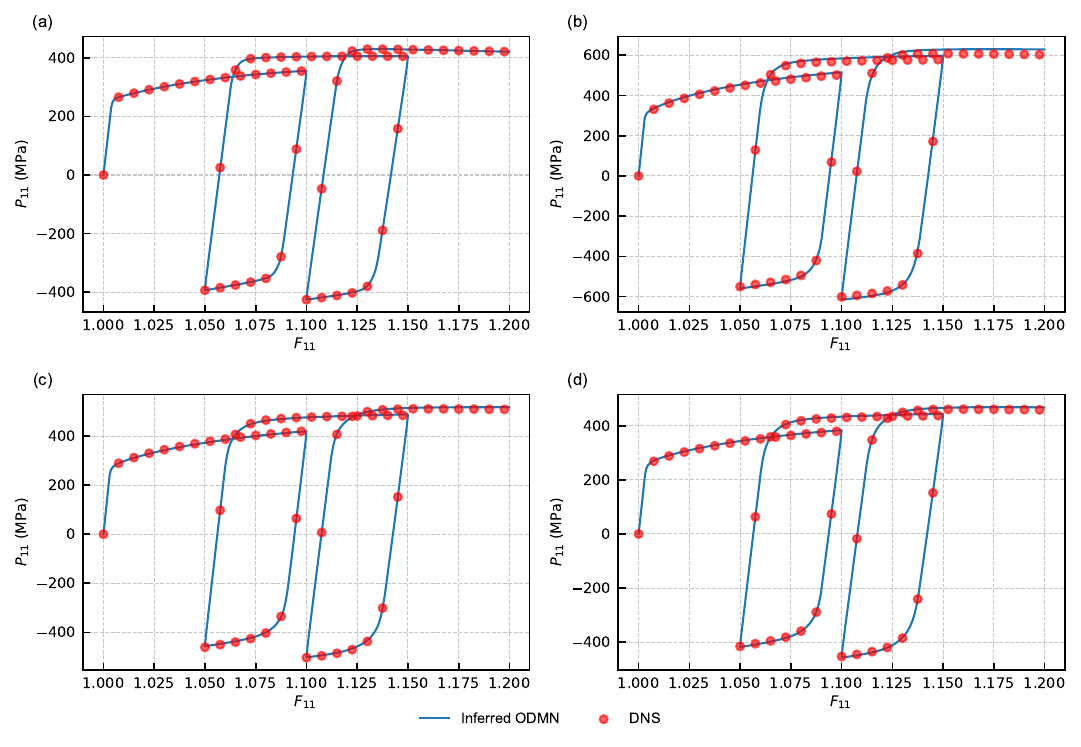}
    \caption{Stress--strain curves under cyclic loading for (a) S1, (b) S2, (c) W1, and (d) W2 at a strain rate of $\dot{F}_{11}=1$. ODMN predictions are compared with DNS results.}
    \label{fig:summary_complex_SScurve}
\end{figure}

Fig.~\ref{fig:summary_complex_SScurve} shows that the stress--strain responses predicted by the inferred ODMNs are in close agreement with the DNS results for all four unseen microstructures. This agreement indicates that the proposed framework can accurately infer microstructure-specific ODMN parameters for previously unseen microstructures. Moreover, the inferred ODMNs retain high predictive accuracy for nonlinear mechanical behavior under cyclic loading. To further quantify the predictive performance, two normalized error metrics are adopted following~\cite{huang2022microstructure}:

\begin{equation}
\text{mean-relative error} =
\frac{ \frac{1}{n} \sum_{i=1}^{n} \left| P_i^{\text{DNS}} - P_i^{\text{ODMN}} \right| }
     { \max_{i=1,\ldots,n} \left| P_i^{\text{DNS}} \right| }
\label{eq:mean_rel_error}
\end{equation}

\begin{equation}
\text{max-relative error} =
\frac{ \max_{i=1,\ldots,n} \left| P_i^{\text{DNS}} - P_i^{\text{ODMN}} \right| }
     { \max_{i=1,\ldots,n} \left| P_i^{\text{DNS}} \right| }
\label{eq:max_rel_error}
\end{equation}

\noindent where $P_i^{\mathrm{DNS}}$ and $P_i^{\mathrm{ODMN}}$ are the $i$-th stress data points from DNS and the inferred ODMN, respectively, and $n$ is the number of data points along the stress--strain curve.

The quantitative results are summarized in Table~\ref{tab:ss_curve_error_metrics_cyclic}. Across all RVEs, the mean relative error remains below 2\%, indicating that the inferred ODMN accurately captures the nonlinear response across various unseen microstructures. The maximum relative error, observed for S2, is 4.32\%, which still represents only a minor deviation from the DNS result. These observations demonstrate the robustness of the proposed framework in predicting nonlinear mechanical responses across diverse microstructures. 

These results further confirm that the learned microstructure-to-parameter mapping generalizes across both unseen microstructures and loading conditions.

\begin{table}[htbp]
\centering
\renewcommand{\arraystretch}{1.3}
\caption{Relative stress prediction errors of inferred ODMNs under cyclic loading, compared against DNS results.}
\begin{tabular}{lcccc}
    \hline
    RVE & S1 & S2 & W1 & W2 \\
    \hline
    mean-relative error (\%) & 0.08 & 1.98 & 0.50 & 0.74 \\
    max-relative error (\%)  & 0.50 & 4.32 & 1.88 & 2.04 \\
    \hline
\end{tabular}
\label{tab:ss_curve_error_metrics_cyclic}
\end{table}


\subsubsection{Cyclic loading: texture evolution prediction}

\begin{figure}[htbp]
    \centering
    \includegraphics[width=1\linewidth]{ 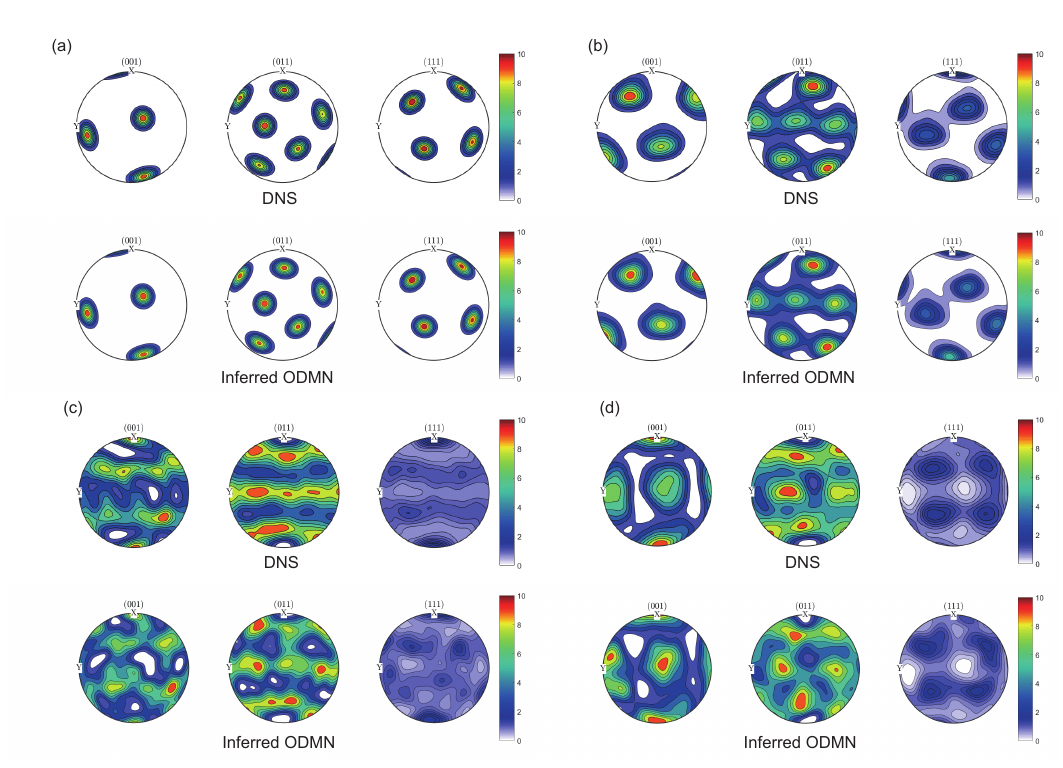}
    \caption{Pole figures after uniaxial cyclic loading with an applied deformation of $F_{11}=1.2$ for (a) S1, (b) S2, (c) W1, and (d) W2. ODMN predictions are compared with DNS results.}
    \label{fig:tensile_texture}
\end{figure}

The capability of the inferred ODMN to predict texture evolution is examined at a deformation of $F_{11}=1.2$. The resulting pole figures in Fig.~\ref{fig:tensile_texture} show close agreement between ODMN and DNS across all RVEs. To quantify this agreement, the normalized texture index of the DODF, $\hat{T}^d$, is used to measure the deviation between the ODFs predicted by ODMN and those obtained from DNS. The corresponding values are summarized in Table~\ref{tab:texture_index_complex}. The results confirm that the inferred ODMN captures the evolution of crystallographic texture without retraining, particularly for strongly textured RVEs. Larger deviations are observed in weakly textured cases, which may be related to the broader orientation distributions in such microstructures and the resulting difficulty of representing them with a compact set of orientations.

\begin{table}[htbp]
\centering
\caption{Normalized texture index of the DODF $\hat{T}^d$ quantifying the difference in ODF between the inferred ODMN and DNS results under uniaxial loading--unloading with $F_{11}=1.2$ and a strain rate of $\dot{F}_{11}=1$.}
\begin{tabular}{lc}
\hline
Microstructure & $\hat{T}^d$ \\
\hline
S1 & 0.0004 \\
S2 & 0.0301 \\
W1 & 0.1176 \\
W2 & 0.1024 \\
\hline
\end{tabular}
\label{tab:texture_index_complex}
\end{table}

\subsubsection{Shear loading: mechanical prediction}

\begin{figure}[htbp]
    \centering
    \includegraphics[width=1\linewidth]{ 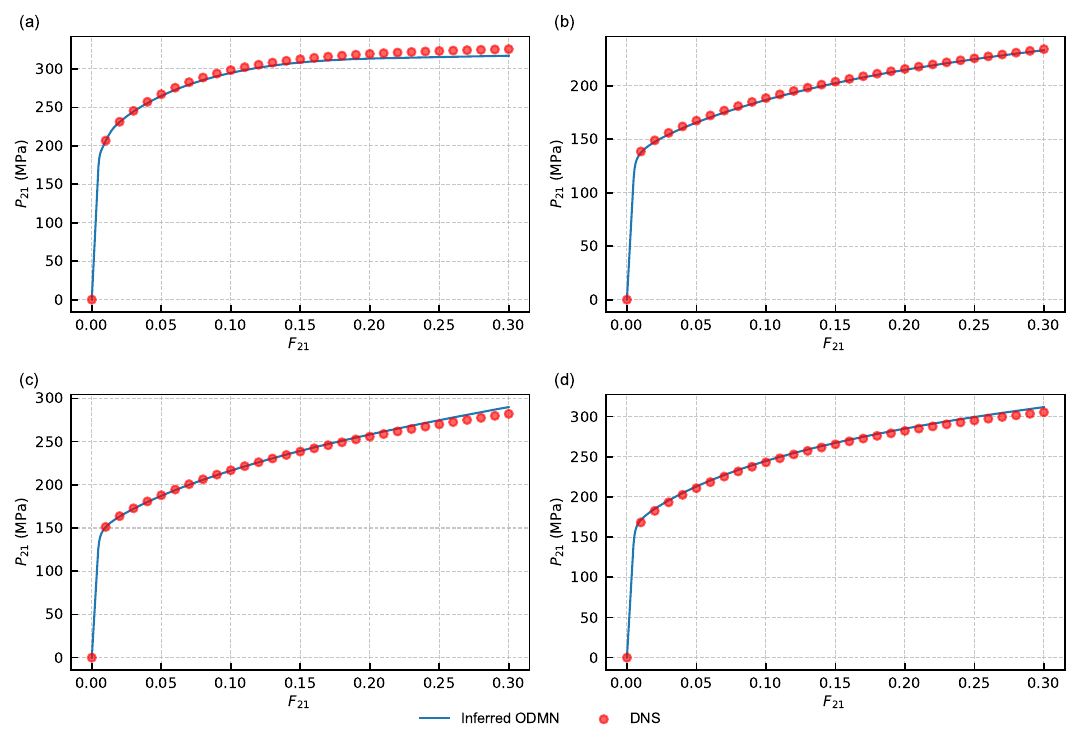}
    \caption{Stress--strain curves under simple shear loading for (a) S1, (b) S2, (c) W1, and (d) W2 at a strain rate of $\dot{F}_{21}=1$. ODMN predictions are compared with DNS results.}
    \label{fig:summary_shear_SScurve}
\end{figure}

The predictive capability of the inferred ODMN is further examined under simple shear loading. As shown in Fig.~\ref{fig:summary_shear_SScurve}, the predicted stress--strain responses again exhibit close agreement with DNS results across all four RVEs. The quantitative results in Table~\ref{tab:ss_curve_error_metrics_shear} show that the mean-relative error remains below 1.5\% and the maximum relative error does not exceed 2.8\%. These results further confirm that the inferred ODMN generalizes across loading conditions, maintaining high predictive accuracy under both cyclic and shear loading.

\begin{table}[htbp]
\centering
\renewcommand{\arraystretch}{1.3}
\caption{Relative stress prediction errors of inferred ODMNs under simple shear loading, compared against DNS results.}
\begin{tabular}{lcccc}
    \hline
    RVE & S1 & S2 & W1 & W2 \\
    \hline
    mean-relative error (\%) & 1.44 & 0.52 & 0.76 & 0.93 \\
    max-relative error (\%)  & 2.63 & 0.77 & 2.78 & 2.16 \\
    \hline
\end{tabular}
\label{tab:ss_curve_error_metrics_shear}
\end{table}

\subsubsection{Shear loading: texture evolution prediction}

\begin{figure}[htbp]
    \centering
    \includegraphics[width=1\linewidth]{ 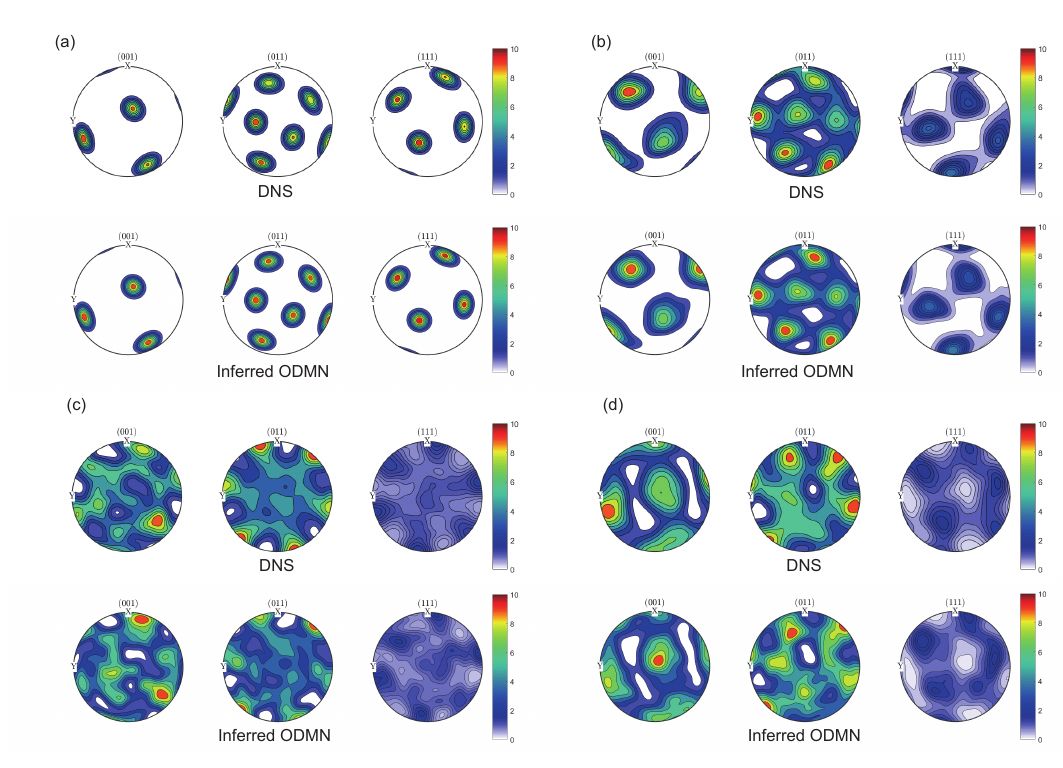}
    \caption{Pole figures after simple shear loading with an applied deformation of $F_{21}=0.3$ for (a) S1, (b) S2, (c) W1, and (d) W2. ODMN predictions are compared with DNS results.}
    \label{fig:shear_texture}
\end{figure}

Texture evolution under simple shear deformation is evaluated at $F_{21}=0.3$ with a strain rate of $\dot{F}_{21}=1$. The pole figures in Fig.~\ref{fig:shear_texture} show good agreement between the DNS and ODMN predictions, indicating that the inferred ODMN captures the evolution of crystallographic texture under shear loading.

The corresponding $\hat{T}^d$ values are summarized in Table~\ref{tab:texture_index_shear}. Consistent with the pole-figure comparison, the discrepancy remains low for strongly textured RVEs but becomes larger for weakly textured cases. This trend reflects the broader and more nearly random orientation distributions of weak textures, which are more difficult to approximate using a compact reduced representation, as also observed in reduced-order electron backscatter diffraction (EBSD) texture representation~\cite{wanni2024machine}. Nevertheless, the inferred ODMN still captures the overall texture evolution with satisfactory accuracy under shear loading.

\begin{table}[htbp]
\centering
\caption{Normalized texture index of the DODF $\hat{T}^d$ quantifying the difference in ODF between the inferred ODMN and DNS results under simple shear deformation with $F_{21}=0.3$ and strain rate $\dot{F}_{21}=1$.}
\begin{tabular}{lc}
\hline
Microstructure & $\hat{T}^d$ \\
\hline
S1 & 0.0101 \\
S2 & 0.0363 \\
W1 & 0.0927 \\
W2 & 0.0911 \\
\hline
\end{tabular}
\label{tab:texture_index_shear}
\end{table}

\subsubsection{Summary of online prediction results}

Overall, the results demonstrate that the TACS--GNN--ODMN framework successfully generalizes the learned microstructure-to-parameter mapping to unseen microstructures under different loading conditions. The framework achieves high accuracy in stress--strain prediction while consistently capturing texture evolution in a physically meaningful manner. Although larger deviations are observed in weakly textured cases, this may reflect the limited expressiveness of the reduced representation. Nevertheless, the framework still provides accurate homogenized mechanical predictions while capturing the main trend of texture evolution.

\subsection{Investigating the Hyperparameter $N$ in the TACS--GNN--ODMN Framework}

\begin{figure}[htbp]
    \centering
    \includegraphics[width=0.85\linewidth]{ 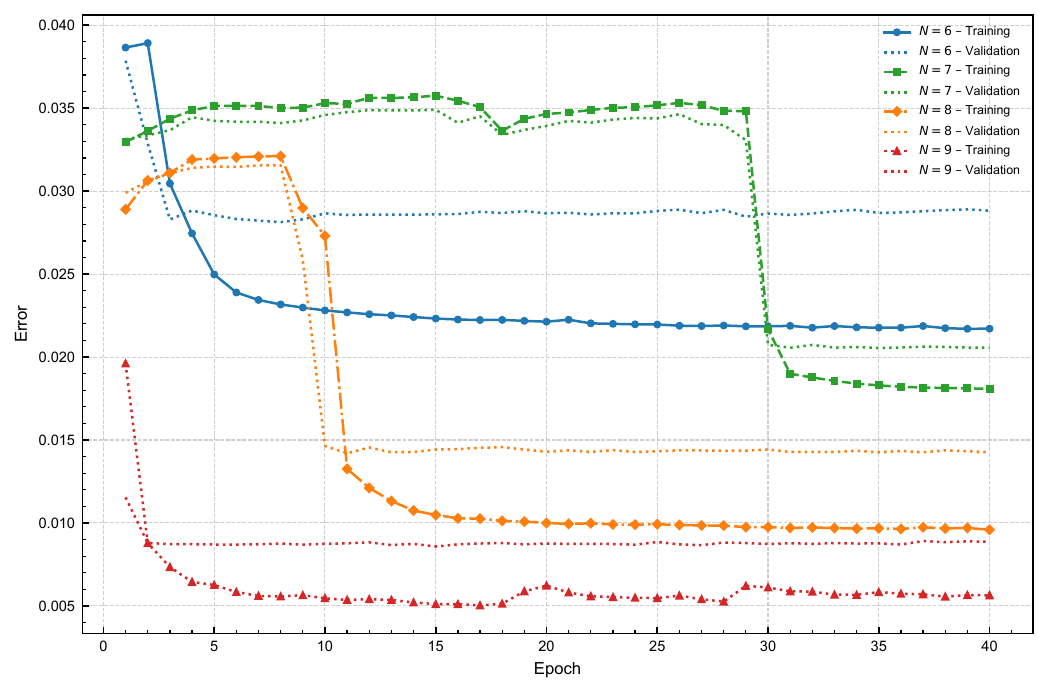}
    \caption{Training and validation error curves of the TACS--GNN--ODMN framework for different values of the hyperparameter $N$.}
    \label{fig:N_depth_curve}
\end{figure}

To investigate the effect of the hyperparameter $N$ on model performance, an ablation study is conducted with $N = 6, 7, 8$, and $9$. For each configuration, the TACS--GNN--ODMN model is trained and validated on the same dataset to ensure a consistent comparison. The corresponding training and validation error curves are presented in Fig.~\ref{fig:N_depth_curve}.

The results show that the validation error decreases monotonically as $N$ increases, indicating that the learned microstructure-to-parameter mapping benefits from the increased representational capacity of ODMN. In particular, when $N = 9$, the validation error falls below 1\%, suggesting that a deeper ODMN provides a more expressive and accurate representation of the underlying microstructure.

From a representational perspective, $N$ controls the resolution of the reduced microstructural description in the ODMN. Smaller values of $N$ yield more compact models with lower computational cost~\cite{Wei01}, whereas larger values provide finer representation of texture and interaction topology at the expense of efficiency. This trade-off highlights the flexibility of the proposed framework in balancing predictive accuracy and computational cost across different application scenarios.

\subsection{Interpretation of the Inferred ODMN as a Reduced Microstructural Representation}

The inferred ODMN may be interpreted as a reduced yet representative description of the underlying microstructure. In particular, following the analogous unit-cell interpretation introduced for DMNs in composite systems~\cite{shin2023deep}, the inferred ODMN may be viewed as an analogous unit cell, in which the RVE is partitioned into $2^N$ subdomains associated with the material nodes. The volume fraction of each subdomain is given by $\frac{W^i}{\sum_{\forall i} W^i}$, while the crystallographic orientation of each subdomain is specified by the Tait--Bryan angles $\alpha^i, \beta^i, \gamma^i$.

The interface normals between subdomains are determined by the stress-equilibrium directions of the material network, denoted by $\vec{\mathbf{N}}^{l}_{p}(\theta^{l}_{p}, \phi^{l}_{p})$:
\begin{equation}\label{eq:direction_vector}
\vec{\mathbf{N}}^{l}_{p} =
\begin{bmatrix}
\cos(2\pi\phi^{l}_{p}) \sin(\pi \theta^{l}_{p}) \\
\sin(2\pi\phi^{l}_{p}) \sin(\pi \theta^{l}_{p}) \\
\cos(\pi \theta^{l}_{p})
\end{bmatrix}.
\end{equation}

\begin{figure}[htbp]
    \centering
    \includegraphics[width=0.85\linewidth]{ 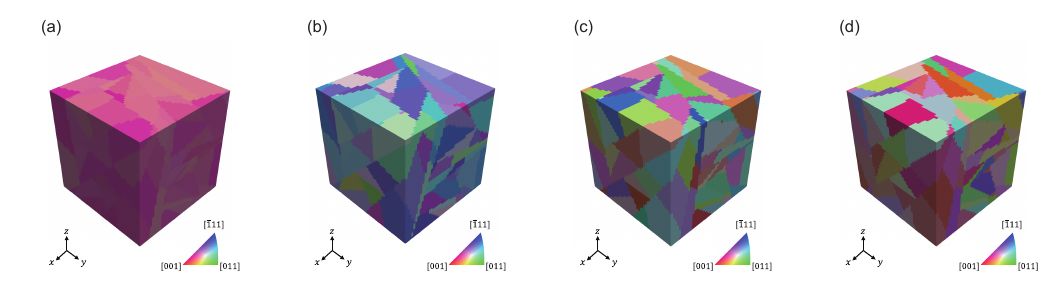}
    \caption{Reconstructed analogous unit cells for the unseen RVEs: (a) S1, (b) S2, (c) W1, and (d) W2.}
    \label{fig:AUCs}
\end{figure}

For the four unseen RVEs, the inferred ODMNs are reconstructed into analogous unit cells, as shown in Fig.~\ref{fig:AUCs}.

The binary-tree partition defining each analogous unit cell is geometrically non-periodic, since every subdomain is obtained by recursively bisecting its parent with a planar interface specified only by a normal direction and a volume fraction. This non-periodicity concerns the geometric partition alone: the homogenization still enforces a periodic kinematic boundary condition on the outer boundary of the cell. Under this condition, the displacement fluctuation field cancels pairwise across both the outer faces and the shared internal interfaces, so that consistency of the volume-averaged deformation gradient is preserved for any internal partition.

\begin{figure}[htbp]
    \centering
    \includegraphics[width=1\linewidth]{ 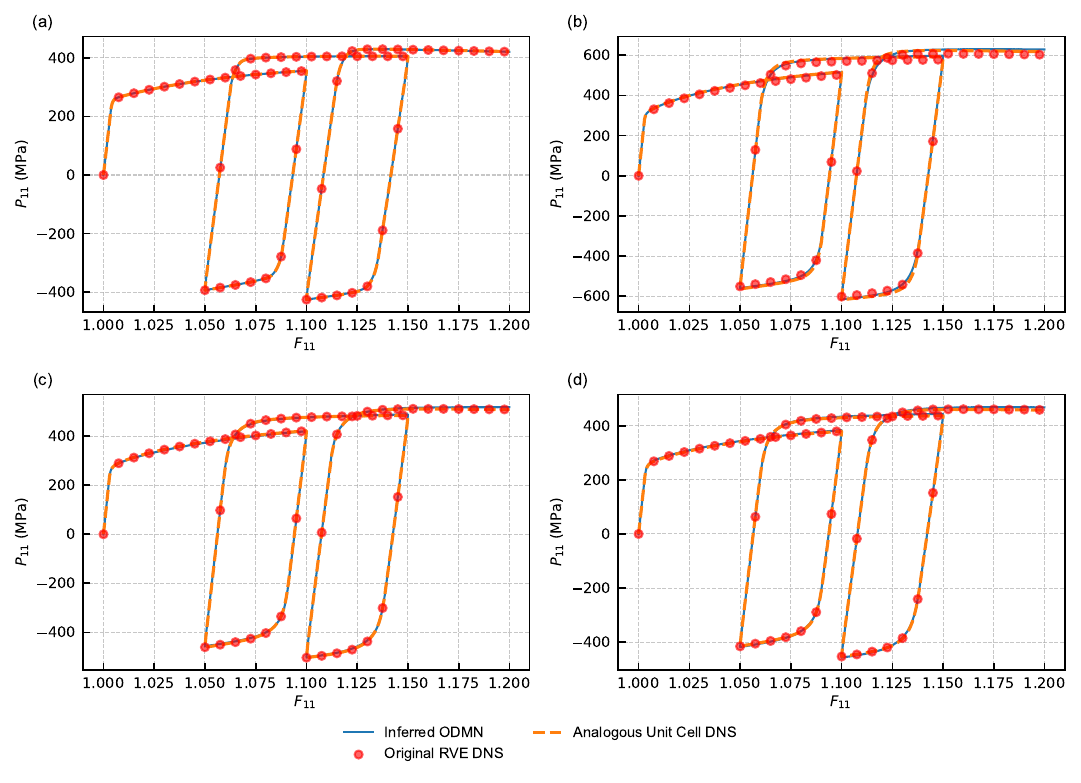}
    \caption{Stress--strain curves under cyclic loading for (a) S1, (b) S2, (c) W1, and (d) W2 at a strain rate of $\dot{F}_{11}=1$. Comparisons among original RVE DNS, inferred ODMN, and analogous unit-cell DNS.}
    \label{fig:summary_complex_SScurve_unitCell}
\end{figure}

Each analogous unit cell is discretized on a $45 \times 45 \times 45$ voxel grid and analyzed using the DAMASK--FFT solver under cyclic loading. The resulting stress--strain responses at $F_{11}=1.2$ are compared with both the DNS results of the original RVE and the corresponding ODMN predictions, as shown in Fig.~\ref{fig:summary_complex_SScurve_unitCell}.

The close agreement among these responses indicates that the inferred ODMN captures not only the homogenized behavior but also the essential microstructural features governing it. This result provides strong evidence that the learned parameter set constitutes a physically meaningful reduced representation of the original microstructure, rather than merely a numerical fit. This observation further supports the capability of the proposed framework to preserve essential microstructural features governing the macroscopic response.

\begin{figure}[htbp]
    \centering
    \includegraphics[width=1\textwidth]{ 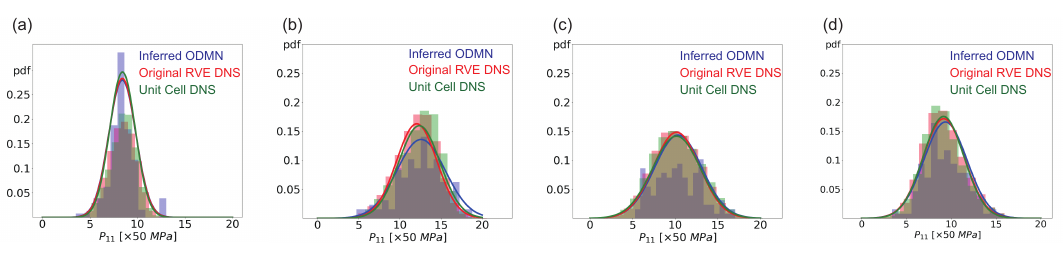}
    \caption{Comparison of local stress distributions predicted by ODMN, original RVE DNS, and analogous unit-cell DNS under uniaxial loading at $F_{11}=1.2$ with $\dot{F}_{11}=1$.}
    \label{fig:stress_dist}
\end{figure}

Beyond the macroscopic response, the local stress distributions are further examined in Fig.~\ref{fig:stress_dist} and Table~\ref{tab:comparison_stress}. Although the reconstructed unit cells differ geometrically from the original RVEs, their statistical stress characteristics remain in close agreement across all approaches.

\begin{table}[htbp]
\centering
\caption{Comparison of local stress distributions ($P_{11}$, in MPa) among ODMN, original RVE DNS, and analogous unit-cell DNS.}
\begin{tabular}{lccc}
\hline
 & ODMN & Original RVE & Analogous unit cell \\
\hline
S1 & 420.41$\pm$71.39 & 420.58$\pm$70.54 & 421$\pm$67.42 \\
S2 & 628.63$\pm$146.82 & 602.34$\pm$122.53 & 617$\pm$125.01 \\
W1 & 517.79$\pm$138.98 & 508.19$\pm$134.59 & 507$\pm$140.83 \\
W2 & 467.79$\pm$120.31 & 458.37$\pm$116.22 & 457.23$\pm$113.74 \\
\hline
\end{tabular}
\label{tab:comparison_stress}
\end{table}

This consistency indicates that the inferred ODMN preserves not only the macroscopic response but also the statistical characteristics of the microscale stress field. Overall, these results demonstrate that the TACS--GNN--ODMN framework provides a physically interpretable and computationally efficient surrogate representation that captures the essential microstructural physics of polycrystalline materials.

\subsection{Computational cost}

The computational efficiency of the proposed TACS--GNN--ODMN framework is evaluated by comparing its inference cost with full-field DNS simulations. The model is trained on an NVIDIA H100 GPU (80\,GB HBM3) with a batch size of 512 for 100 epochs, requiring a total of 30\,h\,40\,min\,30\,s. Once trained, online prediction is performed on Intel\textsuperscript{®} Xeon\textsuperscript{®} Platinum 8480 (2.0\,GHz) central processing units (CPUs).

Table~\ref{tab:computation_time_comparison} summarizes the wall-clock time required by the inferred ODMN and DAMASK--FFT simulations under cyclic and shear loading conditions. The results show that the ODMN achieves a speed-up of more than two orders of magnitude, ranging from approximately $224\times$ to $281\times$ across all cases.

The computational efficiency arises mainly because the inferred ODMN requires constitutive updates only at $2^N$ subdomains during online prediction, rather than across the full-field discretization used in DNS. The resulting reduction in computational complexity enables efficient prediction of the homogenized mechanical response.

These results demonstrate that the proposed framework not only preserves predictive accuracy but also significantly reduces computational cost, making it well-suited for large-scale simulations and real-time applications.

\begin{table}[htbp]
\centering
\caption{Wall time (in CPU-seconds) required for online predictions using the inferred ODMN model and full-field DAMASK FFT simulations. Speed-up is computed as the ratio $T_{\textit{DNS}} / T_{\textit{ODMN}}$.}
\label{tab:computation_time_comparison}
\renewcommand{\arraystretch}{1.2}
\setlength{\tabcolsep}{9pt}
\begin{tabular}{lcccccc}
\toprule
\multirow{2}{*}{RVE} 
& \multicolumn{2}{c}{Inferred ODMN} 
& \multicolumn{2}{c}{DAMASK-FFT} 
& \multicolumn{2}{c}{Speed-up} \\
\cmidrule(r){2-3} \cmidrule(r){4-5} \cmidrule(r){6-7}
& $T_{\textit{Cyclic}}$ & $T_{\textit{Shear}}$ 
& $T_{\textit{Cyclic}}$ & $T_{\textit{Shear}}$ 
& Cyclic & Shear \\
\midrule
S1 & 3031 & 2149 &   782320    &  604464     & $258.1\times$ & $281.3\times$ \\
S2 & 4045 & 2179 &   906528    &    608160   & $224.1\times$ & $279.1\times$ \\
W1 & 3717 & 2373 &   875952    &   613648    & $235.7\times$ & $258.6\times$ \\
W2 & 3884 & 2327 &    871472   &   615776    & $224.4\times$ & $264.6\times$ \\
\bottomrule
\end{tabular}
\end{table}

\section{Conclusions}\label{sec4}

This work presents a computational framework for parametric multiscale surrogate modeling of polycrystalline materials based on the TACS--GNN--ODMN framework. The proposed approach establishes a microstructure-to-parameter mapping, enabling the construction of fully parameterized ODMNs for previously unseen microstructures without retraining.

Within this framework, the TACS module provides a reduced yet statistically consistent representation of crystallographic texture through adaptive sampling of the ODF, while the GNN acts as a parameter-inference operator that predicts micromechanical equilibrium parameters from grain-level interaction graphs. This decomposition enables efficient parameterization of ODMNs while preserving the physical interpretability and computational efficiency of the original framework.

The proposed approach is validated through multiple case studies involving diverse texture types and loading conditions. The results show that the inferred ODMNs accurately reproduce nonlinear mechanical responses and crystallographic texture evolution observed in full-field DNS. In addition, the reconstructed analogous unit cells confirm that the learned parameter set constitutes a physically meaningful reduced representation of the underlying microstructure rather than merely a data-driven fit.

Overall, the proposed framework provides an efficient and physically consistent computational strategy for multiscale modeling of polycrystalline materials. The resulting surrogate model achieves a substantial reduction in computational cost while maintaining high predictive accuracy across previously unseen microstructures and loading conditions.

Future work will focus on extending the framework to broader multiscale modeling pipelines, particularly for establishing structure--process--property linkages at the component scale. Because the inferred ODMN supplies both the homogenized stress and the consistent tangent at each macroscopic material point, it can serve directly as a constitutive driver in structural-scale finite-element analysis. Deploying the framework to system-level problems on the order of $10^{6}$ elements, such as sheet-forming simulations, and validating the predictions against experiments are primary directions of ongoing work. Such extensions may enable efficient exploration of microstructural design spaces and support inverse materials design toward optimized microstructures with targeted performance.

\section*{Acknowledgements}
This work is supported by the National Science and Technology Council, Taiwan, under Grant 111-2221-E-002-054-MY3, 112-2221-E-007-028, and 114-2221-E-002-010-MY3. We are grateful for the computational resources and support from the National Center for Research on Earthquake Engineering (NCREE), NIAR, Taiwan, NTUCE-NCREE Joint Artificial Intelligence Research Center, and the National Center of High-performance Computing (NCHC). 

\section*{Declarations}
The authors declare no competing interests.

\section*{Data Availability}
The datasets and trained models used in this study are available upon request from the corresponding author.

\section*{Author Contributions}
Ting-Ju Wei: Writing – original draft, Methodology, Investigation, Formal analysis, Software, Visualization, Data curation, Conceptualization. Tung-Huan Su: Writing – review \& editing, Methodology, Formal analysis, Supervision, Investigation, Conceptualization. Chuin-Shan Chen: Writing – review \& editing, Supervision, Resources, Project administration, Methodology, Investigation, Funding acquisition, Formal analysis, Conceptualization.

\appendix
\section{Phenomenological Crystal Plasticity Model}
\label{appendixA}

In this work, the phenomenological crystal plasticity framework is adopted as the local constitutive law to capture the material response. The implementation follows the formulation available in DAMASK and neglects the effect of deformation twinning as in ~\cite{roters2019damask}. 
The plastic velocity gradient $\mathbf{L}_p$ is expressed as the accumulated contribution of all slip systems $\alpha$:

\begin{equation}
    \mathbf{L}_p = \sum_{\alpha} \dot{\gamma}^\alpha (\mathbf{s}^\alpha_s \otimes \mathbf{n}^\alpha_s)
\end{equation}
\noindent where $\mathbf{s}^\alpha_s$ denotes the unit vector of the slip direction, $\mathbf{n}^\alpha_s$ the slip plane normal, and $\dot{\gamma}^\alpha$ the shear rate of slip system $\alpha$.  

The evolution of slip resistance $\xi^\alpha$ is modeled as a hardening law that governs its transition from the initial value $\xi_0^\alpha$ toward the saturation value $\xi_{\infty}^\alpha$:

\begin{equation}
    \dot{\xi}^\alpha = h_0^{\text{s-s}}(1+h_{\text{int}}^\alpha) \times\sum_{\alpha'}^{N_s}\left | \dot{\gamma}^{\alpha'} \right |{\left | 1-\frac{\xi ^{\alpha'}}{\xi ^{\alpha'}_\infty} \right |}^{a}\text{sgn}\left(1-\frac{\xi ^{\alpha'}}{\xi ^{\alpha'}_\infty}\right)h^{\alpha\alpha'}
\end{equation}

The slip rate $\dot{\gamma}^\alpha$ is defined in terms of the resolved shear stress $\tau^\alpha$ and the slip resistance $\xi^\alpha$ as

\begin{equation}
    \dot{\gamma}^\alpha = \dot{\gamma}^\alpha_0 {\left | \frac{\tau^\alpha}{\xi^\alpha} \right |}^n \text{sgn}(\tau^\alpha)
\end{equation}

The resolved shear stress $\tau^\alpha$ is obtained from Schmid’s law using the Mandel stress $\mathbf{M}^p$:

\begin{equation}
    \tau^\alpha = \mathbf{M}^p \cdot (\mathbf{s}^\alpha \otimes \mathbf{n}^\alpha)
\end{equation}


\section{Generalized Hooke’s law}
\label{appendixB}

The elastic response of the material is represented through the generalized form of Hooke’s law. 
It establishes the relation between the second Piola–Kirchhoff stress tensor $\mathbf{S}$ and the Green–Lagrange strain tensor $\mathbf{E}$ via the fourth-order stiffness tensor $\mathbb{C}$:

\begin{equation}
    \mathbf{S} = \mathbb{C} : \mathbf{E}
\end{equation}

where
\begin{itemize}
    \item $\mathbf{S}$ is the second Piola–Kirchhoff stress tensor, measured in the reference configuration;
    \item $\mathbb{C}$ denotes the fourth-order elastic stiffness tensor, characterizing the elastic constants of the material;
    \item $\mathbf{E}$ is the Green–Lagrange strain tensor.
\end{itemize}

The Green--Lagrange strain tensor $\mathbf{E}$ is defined directly in terms of the elastic deformation gradient, $\mathbf{F}^e$, as:

\begin{equation}
    \mathbf{E} = \frac{1}{2} \left( (\mathbf{F}^e)^{T} \mathbf{F}^e - \mathbb{I} \right)
\end{equation}
with
\begin{itemize}
    \item $\mathbf{F}^e$ representing the elastic part of the deformation gradient, i.e., the reversible deformation,
    \item $\mathbb{I}$ being the second-order identity tensor.
\end{itemize}

\bibliographystyle{unsrt}  

\bibliography{references}

@article{ZHOU201619,
title = {Experiment and crystal plasticity analysis on plastic deformation of AZ31B Mg alloy sheet under intermediate temperatures: How deformation mechanisms evolve},
journal = {International Journal of Plasticity},
volume = {79},
pages = {19-47},
year = {2016},
issn = {0749-6419},
doi = {https://doi.org/10.1016/j.ijplas.2015.12.006},
author = {Guowei Zhou and Mukesh K. Jain and Peidong Wu and Yichuan Shao and Dayong Li and Yinghong Peng},
}

@article{tjahjanto2015multiscale,
doi = {https://doi.org/10.1088/0965-0393/23/4/045005},
year = {2015},
month = {apr},
publisher = {IOP Publishing},
volume = {23},
number = {4},
pages = {045005},
author = {Tjahjanto, D D and Eisenlohr, P and Roters, F},
title = {Multiscale deep drawing analysis of dual-phase steels using grain cluster-based RGC scheme},
journal = {Modelling and Simulation in Materials Science and Engineering},
}

@article{eisenlohr2013spectral,
title = {A spectral method solution to crystal elasto-viscoplasticity at finite strains},
journal = {International Journal of Plasticity},
volume = {46},
pages = {37-53},
year = {2013},
issn = {0749-6419},
doi = {https://doi.org/10.1016/j.ijplas.2012.09.012},
author = {P. Eisenlohr and M. Diehl and R.A. Lebensohn and F. Roters}
}

@article{ shanthraj2015numerically,
Author = {Shanthraj, P. and Eisenlohr, P. and Diehl, M. and Roters, F.},
Title = {Numerically robust spectral methods for crystal plasticity simulations of heterogeneous materials},
Journal = {International Journal of Plasticity},
Year = {2015},
Volume = {66},
Number = {SI},
Pages = {31-45},
Month = {MAR},
DOI = {https://doi.org/10.1016/j.ijplas.2014.02.006},
ISSN = {0749-6419},
EISSN = {1879-2154},
ResearcherID-Numbers = {Diehl, Martin/A-2831-2016
   Eisenlohr, Philip/E-6866-2010
   },
ORCID-Numbers = {Shanthraj, Pratheek/0000-0002-6324-0306
   Diehl, Martin/0000-0002-3738-7363
   Eisenlohr, Philip/0000-0002-8220-5995
   Roters, Franz/0000-0002-9098-9566},
Unique-ID = {WOS:000349879500003},
}

@article{ lebensohn2020spectral,
Author = {Lebensohn, Ricardo A. and Rollett, Anthony D.},
Title = {Spectral methods for full-field micromechanical modelling of
   polycrystalline materials},
Journal = {Computational materials science},
Year = {2020},
Volume = {173},
Month = {FEB 15},
DOI = {https://doi.org/10.1016/j.commatsci.2019.109336},
Article-Number = {109336},
ISSN = {0927-0256},
EISSN = {1879-0801},
ResearcherID-Numbers = {Rollett, Anthony/A-4096-2012
   Lebensohn, Ricardo/A-2494-2008},
ORCID-Numbers = {Lebensohn, Ricardo/0000-0002-3152-9105},
Unique-ID = {WOS:000506172700006},
}

@article{anand2004single,
title = {Single-crystal elasto-viscoplasticity: application to texture evolution in polycrystalline metals at large strains},
journal = {Computer Methods in Applied Mechanics and Engineering},
volume = {193},
number = {48},
pages = {5359-5383},
year = {2004},
note = {Advances in Computational Plasticity},
issn = {0045-7825},
doi = {https://doi.org/10.1016/j.cma.2003.12.068},
author = {Lallit Anand}
}

@article{ardeljan2014dislocation,
title = {A dislocation density based crystal plasticity finite element model: Application to a two-phase polycrystalline HCP/BCC composites},
journal = {Journal of the Mechanics and Physics of Solids},
volume = {66},
pages = {16-31},
year = {2014},
issn = {0022-5096},
doi = {https://doi.org/10.1016/j.jmps.2014.01.006},
author = {Milan Ardeljan and Irene J. Beyerlein and Marko Knezevic}
}

@article{knezevic2010deformation,
title = {Deformation twinning in {AZ31}: Influence on strain hardening and texture evolution},
journal = {Acta Materialia},
volume = {58},
number = {19},
pages = {6230-6242},
year = {2010},
issn = {1359-6454},
doi = {https://doi.org/10.1016/j.actamat.2010.07.041},
author = {Marko Knezevic and Amanda Levinson and Ryan Harris and Raja K. Mishra and Roger D. Doherty and Surya R. Kalidindi}
}

@article{IBRAGIMOVA2022103374,
title = {A convolutional neural network based crystal plasticity finite element framework to predict localised deformation in metals},
journal = {International Journal of Plasticity},
volume = {157},
pages = {103374},
year = {2022},
issn = {0749-6419},
doi = {https://doi.org/10.1016/j.ijplas.2022.103374},
url = {https://www.sciencedirect.com/science/article/pii/S0749641922001541},
author = {Olga Ibragimova and Abhijit Brahme and Waqas Muhammad and Daniel Connolly and Julie Lévesque and Kaan Inal}
}

@article{hu2024temporal,
title = {A temporal graph neural network for cross-scale modelling of polycrystals considering microstructure interaction},
journal = {International Journal of Plasticity},
volume = {179},
pages = {104017},
year = {2024},
issn = {0749-6419},
doi = {https://doi.org/10.1016/j.ijplas.2024.104017},
author = {Yuanzhe Hu and Guowei Zhou and Myoung-Gyu Lee and Peidong Wu and Dayong Li}
}

@article{roters2019damask,
title = {{DAMASK} – The Düsseldorf Advanced Material Simulation Kit for modeling multi-physics crystal plasticity, thermal, and damage phenomena from the single crystal up to the component scale},
journal = {Computational Materials Science},
volume = {158},
pages = {420-478},
year = {2019},
issn = {0927-0256},
doi = {https://doi.org/10.1016/j.commatsci.2018.04.030},
author = {F. Roters and M. Diehl and P. Shanthraj and P. Eisenlohr and C. Reuber and S.L. Wong and T. Maiti and A. Ebrahimi and T. Hochrainer and H.-O. Fabritius and S. Nikolov and M. Friák and N. Fujita and N. Grilli and K.G.F. Janssens and N. Jia and P.J.J. Kok and D. Ma and F. Meier and E. Werner and M. Stricker and D. Weygand and D. Raabe},
}

@article{SUN2021102973,
title = {Cross-scale prediction from RVE to component},
journal = {International Journal of Plasticity},
volume = {140},
pages = {102973},
year = {2021},
issn = {0749-6419},
doi = {https://doi.org/10.1016/j.ijplas.2021.102973},
author = {Xinxin Sun and Hongwei Li and Mei Zhan and Junyuan Zhou and Jian Zhang and Jia Gao}
}

@article{IBRAGIMOVA2021103059,
title = {A new ANN based crystal plasticity model for FCC materials and its application to non-monotonic strain paths},
journal = {International Journal of Plasticity},
volume = {144},
pages = {103059},
year = {2021},
issn = {0749-6419},
doi = {https://doi.org/10.1016/j.ijplas.2021.103059},
author = {Olga Ibragimova and Abhijit Brahme and Waqas Muhammad and Julie Lévesque and Kaan Inal}
}

@article{ZHOU2024115861,
title = {A physics-constrained neural network for crystal plasticity modelling of FCC materials},
journal = {Scripta Materialia},
volume = {241},
pages = {115861},
year = {2024},
issn = {1359-6462},
doi = {https://doi.org/10.1016/j.scriptamat.2023.115861},
author = {Guowei Zhou and Yuanzhe Hu and Zizheng Cao and Myoung Gyu Lee and Dayong Li}
}

@article{BONATTI2022104697,
title = {On the importance of self-consistency in recurrent neural network models representing elasto-plastic solids},
journal = {Journal of the Mechanics and Physics of Solids},
volume = {158},
pages = {104697},
year = {2022},
issn = {0022-5096},
doi = {https://doi.org/10.1016/j.jmps.2021.104697},
author = {Colin Bonatti and Dirk Mohr}
}

@article{BONATTI2022103430,
title = {From {CP-FFT} to {CP-RNN}: Recurrent neural network surrogate model of crystal plasticity},
journal = {International Journal of Plasticity},
volume = {158},
pages = {103430},
year = {2022},
issn = {0749-6419},
doi = {https://doi.org/10.1016/j.ijplas.2022.103430},
author = {Colin Bonatti and Bekim Berisha and Dirk Mohr}
}

@article{Wei01,
title = {Orientation-aware interaction-based deep material network in polycrystalline materials modeling},
journal = {Computer Methods in Applied Mechanics and Engineering},
volume = {441},
pages = {117977},
year = {2025},
issn = {0045-7825},
doi = {https://doi.org/10.1016/j.cma.2025.117977},
author = {Ting-Ju Wei and Tung-Huan Su and Chuin-Shan Chen},
}

@article{WeiOverview,
author = {Ting-Ju Wei and Wen-Ning Wan and Chuin-Shan Chen},
title = {Deep Material Network: Overview, Applications and Current Directions},
journal = {Multiscale Science and Engineering},
volume = {8},
pages = {1--19},
doi = {https://doi.org/10.1007/s42493-026-00146-4},
year = {2026}
}

@article{wan2024decoding,
  title={Decoding material networks: exploring performance of deep material network and interaction-based material networks},
  author={Wan, Wen-Ning and Wei, Ting-Ju and Su, Tung-Huan and Chen, Chuin-Shan},
  journal={Journal of Mechanics},
  volume={40},
  pages={796--807},
  year={2024},
  publisher={Oxford University Press},
  doi = {https://doi.org/10.1093/jom/ufae053}
}

@article{LIU201920,
title = {Exploring the 3D architectures of deep material network in data-driven multiscale mechanics},
journal = {Journal of the Mechanics and Physics of Solids},
volume = {127},
pages = {20-46},
year = {2019},
issn = {0022-5096},
doi = {https://doi.org/10.1016/j.jmps.2019.03.004},
author = {Zeliang Liu and C.T. Wu},
}

@article{noels2022micromechanics,
title = {Micromechanics-based material networks revisited from the interaction viewpoint; robust and efficient implementation for multi-phase composites},
journal = {European Journal of Mechanics - A/Solids},
volume = {91},
pages = {104384},
year = {2022},
issn = {0997-7538},
doi = {https://doi.org/10.1016/j.euromechsol.2021.104384},
author = {Van Dung Nguyen and Ludovic Noels},
}

@article{wanni2024machine,
  title={Machine learning enhanced analysis of EBSD data for texture representation},
  author={Wanni, J and Bronkhorst, CA and Thoma, DJ},
  journal={npj Computational Materials},
  volume={10},
  pages={133},
  year={2024},
  doi = {https://doi.org/10.1038/s41524-024-01324-4},
  publisher={Nature Publishing Group UK London}
}

@article{noels2022interaction,
title = {Interaction-based material network: A general framework for (porous) microstructured materials},
journal = {Computer Methods in Applied Mechanics and Engineering},
volume = {389},
pages = {114300},
year = {2022},
issn = {0045-7825},
doi = {https://doi.org/10.1016/j.cma.2021.114300},
author = {Van Dung Nguyen and Ludovic Noels},
}

@article{shin2024deep2,
title = {Deep material network for thermal conductivity problems: Application to woven composites},
journal = {Computer Methods in Applied Mechanics and Engineering},
volume = {431},
pages = {117279},
year = {2024},
issn = {0045-7825},
doi = {https://doi.org/10.1016/j.cma.2024.117279},
author = {Dongil Shin and Peter Jefferson Creveling and Scott Alan Roberts and Rémi Dingreville},
}

@article{jean2024graph,
  title   = {Graph-enhanced deep material network: multiscale materials modeling with microstructural informatics},
  journal = {Computational Mechanics},
  volume  = {75},
  number  = {1},
  pages   = {113--136},
  year    = {2025},
  doi     = {https://doi.org/10.1007/s00466-024-02493-1},
  issn    = {1432-0924},
  author  = {Jean, Jimmy Gaspard and Su, Tung-Huan and Huang, Szu-Jui and Wu, Cheng-Tang and Chen, Chuin-Shan},
}

@article{gajek2020micromechanics,
title = {On the micromechanics of deep material networks},
journal = {Journal of the Mechanics and Physics of Solids},
volume = {142},
pages = {103984},
year = {2020},
issn = {0022-5096},
doi = {https://doi.org/10.1016/j.jmps.2020.103984},
author = {Sebastian Gajek and Matti Schneider and Thomas Böhlke},
}

@article{sterr2025deep,
author = {Sterr, Benedikt and Gajek, Sebastian and Hrymak, Andrew and Schneider, Matti and Böhlke, Thomas},
title = {Deep Material Networks for Fiber Suspensions With Infinite Material Contrast},
journal = {International Journal for Numerical Methods in Engineering},
volume = {126},
number = {7},
pages = {e70014},
doi = {https://doi.org/10.1002/nme.70014},
year = {2025}
}

@article{wei2025foundation,
title = {Foundation model for composite microstructures: Reconstruction, stiffness, and nonlinear behavior prediction},
journal = {Materials \& Design},
volume = {257},
pages = {114397},
year = {2025},
issn = {0264-1275},
doi = {https://doi.org/10.1016/j.matdes.2025.114397},
author = {Ting-Ju Wei and Chuin-Shan Chen},
}

@inproceedings{brody2022how,
  title     = {How Attentive are Graph Attention Networks?},
  author    = {Shaked Brody and Uri Alon and Eran Yahav},
  booktitle = {International Conference on Learning Representations},
  year      = {2022},
  url       = {https://openreview.net/forum?id=F72ximsx7C1}
}

@article{DAI2021117006,
title = {Studying the micromechanical behaviors of a polycrystalline metal by artificial neural networks},
journal = {Acta Materialia},
volume = {214},
pages = {117006},
year = {2021},
issn = {1359-6454},
doi = {https://doi.org/10.1016/j.actamat.2021.117006},
author = {Wei Dai and Huamiao Wang and Qiang Guan and Dayong Li and Yinghong Peng and Carlos N. Tomé},
}

@article{groeber2014dream,
  title={{DREAM. 3D}: a digital representation environment for the analysis of microstructure in 3D},
  author={Groeber, Michael A and Jackson, Michael A},
  journal={Integrating materials and manufacturing innovation},
  volume={3},
  number={1},
  pages={56--72},
  year={2014},
  doi = {https://doi.org/10.1186/2193-9772-3-5},
  publisher={Springer}
}

@article{barrett2019deep,
title = {Deep drawing simulations using the finite element method embedding a multi-level crystal plasticity constitutive law: Experimental verification and sensitivity analysis},
journal = {Computer Methods in Applied Mechanics and Engineering},
volume = {354},
pages = {245-270},
year = {2019},
issn = {0045-7825},
doi = {https://doi.org/10.1016/j.cma.2019.05.035},
author = {Timothy J. Barrett and Marko Knezevic},

}

@misc{damask_documentation,
  title        = {{DAMASK} Documentation: Phenopowerlaw AA6022-T4},
  author       = {{Max-Planck-Institut für Nachhaltige Materialien GmbH}},
  year         = {2024},
  url          = {https://damask-multiphysics.org/documentation/},
  note         = {Accessed: 21 June 2024}
}

@article{huang2022microstructure,
title = {Microstructure-guided deep material network for rapid nonlinear material modeling and uncertainty quantification},
journal = {Computer Methods in Applied Mechanics and Engineering},
volume = {398},
pages = {115197},
year = {2022},
issn = {0045-7825},
doi = {https://doi.org/10.1016/j.cma.2022.115197},
author = {Tianyu Huang and Zeliang Liu and C.T. Wu and Wei Chen},
}

@article{shin2023deep,
  title={Deep material network via a quilting strategy: visualization for explainability and recursive training for improved accuracy},
  author={Shin, Dongil and Alberdi, Ryan and Lebensohn, Ricardo A and Dingreville, R{\'e}mi},
  journal={npj Computational Materials},
  volume={9},
  number={1},
  pages={128},
  year={2023},
  doi = {https://doi.org/10.1038/s41524-023-01085-6},
  publisher={Nature Publishing Group UK London}
}

@article{WEBER2022115384,
title = {Machine learning-enabled self-consistent parametrically-upscaled crystal plasticity model for Ni-based superalloys},
journal = {Computer Methods in Applied Mechanics and Engineering},
volume = {402},
pages = {115384},
year = {2022},
issn = {0045-7825},
doi = {https://doi.org/10.1016/j.cma.2022.115384},
author = {George Weber and Maxwell Pinz and Somnath Ghosh}
}

@article{KNEZEVIC2014239,
title = {Three dimensional predictions of grain scale plasticity and grain boundaries using crystal plasticity finite element models},
journal = {Computer Methods in Applied Mechanics and Engineering},
volume = {277},
pages = {239-259},
year = {2014},
issn = {0045-7825},
doi = {https://doi.org/10.1016/j.cma.2014.05.003},
author = {Marko Knezevic and Borys Drach and Milan Ardeljan and Irene J. Beyerlein}
}

@article{BHATKEELANJESRINIVAS2026118517,
title = {Rapid Offline Training for Deep Material Networks via a displacement-based laminate formulation and a novel sampling technique for a compliance-based fatigue model},
journal = {Computer Methods in Applied Mechanics and Engineering},
volume = {449},
pages = {118517},
year = {2026},
issn = {0045-7825},
doi = {https://doi.org/10.1016/j.cma.2025.118517},
author = {Pavan {Bhat Keelanje Srinivas} and Matthias Kabel and Matti Schneider},
}

@article{FRANCIS2025118329,
title = {Micropolar deep material network},
journal = {Computer Methods in Applied Mechanics and Engineering},
volume = {446},
pages = {118329},
year = {2025},
issn = {0045-7825},
doi = {https://doi.org/10.1016/j.cma.2025.118329},
author = {Noah M. Francis and Dongil Shin and Ricardo A. Lebensohn and Fatemeh Pourahmadian and Rémi Dingreville}
}

@article{WU2026118554,
title = {Convolutional neural network-based mapping of material micro-structures to deep material networks for non-linear mechanical response prediction},
journal = {Computer Methods in Applied Mechanics and Engineering},
volume = {449},
pages = {118554},
year = {2026},
issn = {0045-7825},
doi = {https://doi.org/10.1016/j.cma.2025.118554},
author = {Ling Wu and Ludovic Noels}
}

@article{LI2024116687,
title = {Micromechanics-informed parametric deep material network for physics behavior prediction of heterogeneous materials with a varying morphology},
journal = {Computer Methods in Applied Mechanics and Engineering},
volume = {419},
pages = {116687},
year = {2024},
issn = {0045-7825},
doi = {https://doi.org/10.1016/j.cma.2023.116687},
author = {Tianyi Li}
}

@article{su2022multiscale,
  title   = {Multiscale computational solid mechanics: data and machine learning},
  author  = {Su, Tung-Huan and Huang, Szu-Jui and Jean, Jimmy Gaspard and Chen, Chuin-Shan},
  journal = {Journal of Mechanics},
  volume  = {38},
  pages   = {568--585},
  year    = {2022},
  doi     = {10.1093/jom/ufac037}
}

@article{chou2024structgnn,
  title   = {{StructGNN}: An efficient graph neural network framework for static structural analysis},
  author  = {Chou, Yuan-Tung and Chang, Wei-Tze and Jean, Jimmy Gaspard and Chang, Kai-Hung and Huang, Yin-Nan and Chen, Chuin-Shan},
  journal = {Computers \& Structures},
  volume  = {299},
  pages   = {107385},
  year    = {2024},
  doi     = {10.1016/j.compstruc.2024.107385}
}

\end{document}